# Improving Bias Correction Methods for Daily Rainfall Using a Markov Chain Approach

Danny Parsons* [1,2,3], David Stern [3], Mouhamadou Bamba Sylla [2], James Musyoka [3], John Bagiliko [1,2], Lily Clements [3], John Mupuro [4], Denis Ndanguza [1]

[1] Department of Mathematics, School of Science, College of Science and Technology, University of Rwanda, Kigali P.O. Box 3900, Rwanda

[2] African Institute for Mathematical Sciences (AIMS), AIMS Research and Innovation Centre, KN 3 Rd, Kigali, Kicukiro, Rwanda

[3] IDEMS International, Reading RG2 7AX, UK

[4] Meteorological Services Department, Belvedere, Harare, Zimbabwe

* Corresponding author, danny@aims.ac.za

**Abstract**

Accurate, localised rainfall information is essential for agricultural planning, climate risk assessment, and water resources management. Gridded climate products provide rainfall information over large areas but can lack the accuracy needed at local scales, often requiring bias correction before use in local impact studies. Local intensity scaling (LOCI) and quantile mapping (QM) are two widely used bias correction methods which adjust both rainfall frequency and intensity, but do not account for the temporal structure of daily rainfall. This can lead to biases in the representation of wet and dry spells. This study proposes integrating a two-state first-order Markov chain into existing bias correction methods through state-dependent rain day thresholds and rainfall adjustments, aimed at improving temporal structure. Two implementations of this framework are presented: Markov chain local intensity scaling (MC LOCI) and Markov chain quantile mapping (MC QM). The proposed methods were applied to AgERA5 reanalysis data with rainfall data from five stations in Zimbabwe. Results showed that the Markov chain methods improved the representation of rainfall persistence, onset, and wet and dry spell characteristics compared to LOCI and QM, while maintaining improvements in rain day frequency, mean and total rainfall. Improvements in event timing and daily rainfall amounts were limited. Results from five locations in Zimbabwe demonstrate that the proposed methods could be beneficial for crop simulation, hydrological modelling and other applications requiring accurate rainfall sequencing. Evaluation across additional regions and gridded products would establish the broader applicability of the proposed methods under a range of conditions.

**Keywords:**

Bias correction, Precipitation, Markov chain, Local intensity scaling, Quantile mapping, Downscaling, Reanalysis data

**Acknowledgments**

The authors gratefully acknowledge and thank Rebecca Manzou and the Meteorological Services Department of Zimbabwe for providing station rainfall data and for their valuable support throughout this study.

# 1 Introduction

Access to high quality, localised rainfall information is critical in many areas including agricultural productivity (Amare, McKay, Tarp, & Barrett, 2018), understanding climate change effects (IPCC, 2022), water resources management (He, Sonnenborg, Refsgaard, Vejen, & Jensen, 2013), and climate risk assessment (IPCC, 2022; Sillmann, et al., 2017), particularly in water scarce environments. Many gridded climate products provide timely, long term rainfall information over large areas with high temporal resolution such as global and regional climate models (Eyring, et al., 2016; Rummukainen, 2010; Giorgi & Gutowski Jr, 2015), reanalysis models (Hersbach, et al., 2020; Muñoz-Sabater, et al., 2021; Boogaard, et al., 2020; Ebita, et al., 2011), and satellite based estimates (Funk, et al., 2015; Maidment, et al., 2017; Huffman, et al., 2019). However, these data often exhibit systematic biases in various rainfall characteristics including rainfall occurrence, persistence, and intensity (Teutschbein & Seibert, 2012; Cannon, Sobie, & Murdock, 2015). Biases are partly due to the gridded nature of climate products, which represent large areas, even for high resolution products, and hence limit their direct applicability in localised studies.

Bias correction methods are widely used to reduce systematic biases in gridded rainfall data, making them better suited for localised studies. Bias correction is more complex for daily rainfall than for other climate variables such as temperature, due to the frequency of zero values and spatial and temporal heterogeneity (Soo, et al., 2020). Simple bias correction methods such as power transformation and linear scaling fail to correct discrepancies in rain day frequency (Soo, et al., 2020). Many applications also require the simultaneous representation of multiple rainfall characteristics; for example, occurrence and intensity of rainfall, timing and magnitude of (extreme) events, and spell persistence. This places further demand on bias correction approaches.

Methods have been developed to directly address the overestimation of rain days of gridded data. These include mean based methods such as local intensity scaling (LOCI) (Schmidli, Frei, & Vidale, 2006), and distribution based methods such as quantile mapping (QM) (Themeßl, Gobiet, & Leuprecht, 2011; Piani, et al., 2010; Teutschbein & Seibert, 2012; Panofsky & Brier, 1968; Ines & Hansen, 2006). These methods first correct for bias in rain day frequency by defining a rain day threshold, usually on a seasonal or monthly basis, such that the adjusted rain day frequency matches the rain day frequency of the observation data in the calibration period. This is followed by an adjustment to rainfall amounts above the threshold, either through a scale factor to correct mean rainfall intensity, or a distribution mapping to correct the entire rainfall distribution. However, these methods do not consider the temporal structure of rainfall occurrence and intensity, which may lead to biases, for example, in the representation of wet and dry spells and rainfall intensity dynamics, important characteristics for many applications, such as crop simulation modelling (Timlin, Paff, & Han, 2024).

The temporal structure of daily rainfall has been successfully modelled using Markov chains in various contexts. For example, Stern & Cooper (2011) applied a Markov chain model to daily rainfall occurrence in a case study from Zambia, using long-term rainfall records to derive climate risk metrics such as the probability of dry spells and the onset of the rainy season for agricultural decision making. Two-state first-order Markov chains are often used to model daily rainfall occurrence in stochastic weather generators (Gabriel & Neumann, 1962; Wilks, 1999b; Ailliot, Allard, Monbet, & Naveau, 2015). Liu et al. (2020) used a two-state first-order Markov chain as a separate sequencing stage after a quantile mapping bias correction to improve the temporal representation of rainfall occurrence. However, despite their ability to effectively represent the temporal structure of rainfall, Markov chains have not been incorporated directly into standard bias correction methods, including LOCI and QM.

This study proposes to directly integrate Markov chains into two widely used bias correction methods, LOCI and QM, with the aim of improving the representation of the temporal structure of daily rainfall occurrence and intensity. The proposed methods, Markov chain local intensity scaling (MC LOCI) and Markov chain quantile mapping (MC QM), are designed to improve temporal structure through the application of a rain day threshold for each of the two states of a first-order Markov chain, based on the transition probabilities from the reference data. Rainfall values above the threshold are then adjusted using either a state-dependent scale factor (for MC LOCI) or distribution mapping (for MC QM), again based on the reference data characteristics.

Improved representation of temporal structure could benefit a range of applications that rely on realistic wet and dry spell patterns. For example, more accurate rainfall estimates could support agricultural climate information services in Zimbabwe, where smallholder farmers are highly sensitive to rainfall onset and within-season dry spell variability (Rurinda, et al., 2014).

The proposed methods are evaluated against the original bias correction methods using the European Centre for Medium-Range Weather Forecasts (ECMWF) Agrometeorological reanalysis dataset (AgERA5) (Boogaard, et al., 2020) as the gridded input data and gauge data from five stations in Zimbabwe as the reference dataset. Performance is evaluated across climatology, annual summaries, spell distributions, seasonal structure, and rainfall detection, for both rainfall occurrence and amounts. The aim of this study is to determine whether integrating Markov chains into two standard bias correction methods improves the temporal structure of daily rainfall and the representation of other rainfall characteristics compared with the original bias correction methods, and to identify limitations and directions for future development.

# 2 Study Area

Five sites in Zimbabwe, Southern Africa, are used in this study, as shown in Figure 1. Zimbabwe has a subtropical climate (Mushawemhuka, Fitchett, & Hoogendoorn, 2021) with a distinct rainy season from late October to March in most of the country, and a dry season typically from April to September, characterised by cooler temperatures and very limited rainfall (Mazvimavi, 2010; Mapurisa & Chikodzi, 2014). According to the Köppen climate classification, Zimbabwe is characterised by five climatic zones (Chen & Chen, 2013). The sites used in this study fall within the two largest of these: a hot semi-arid climate, experienced in the majority of the southern and western parts of the country, and a humid subtropical climate, which covers much of the north and eastern regions.

According to Chen & Chen (2013), the four sites in southern Zimbabwe - Plumtree in the southwest, and Masvingo, Chisumbanje and Buffalo Range in the southeast, are in the hot semi-arid zone, characterised by high temperatures and relatively low annual rainfall. Mt Darwin, in northeast Zimbabwe, is in the humid subtropical zone, typically associated with higher annual rainfall and warm temperatures throughout the year.

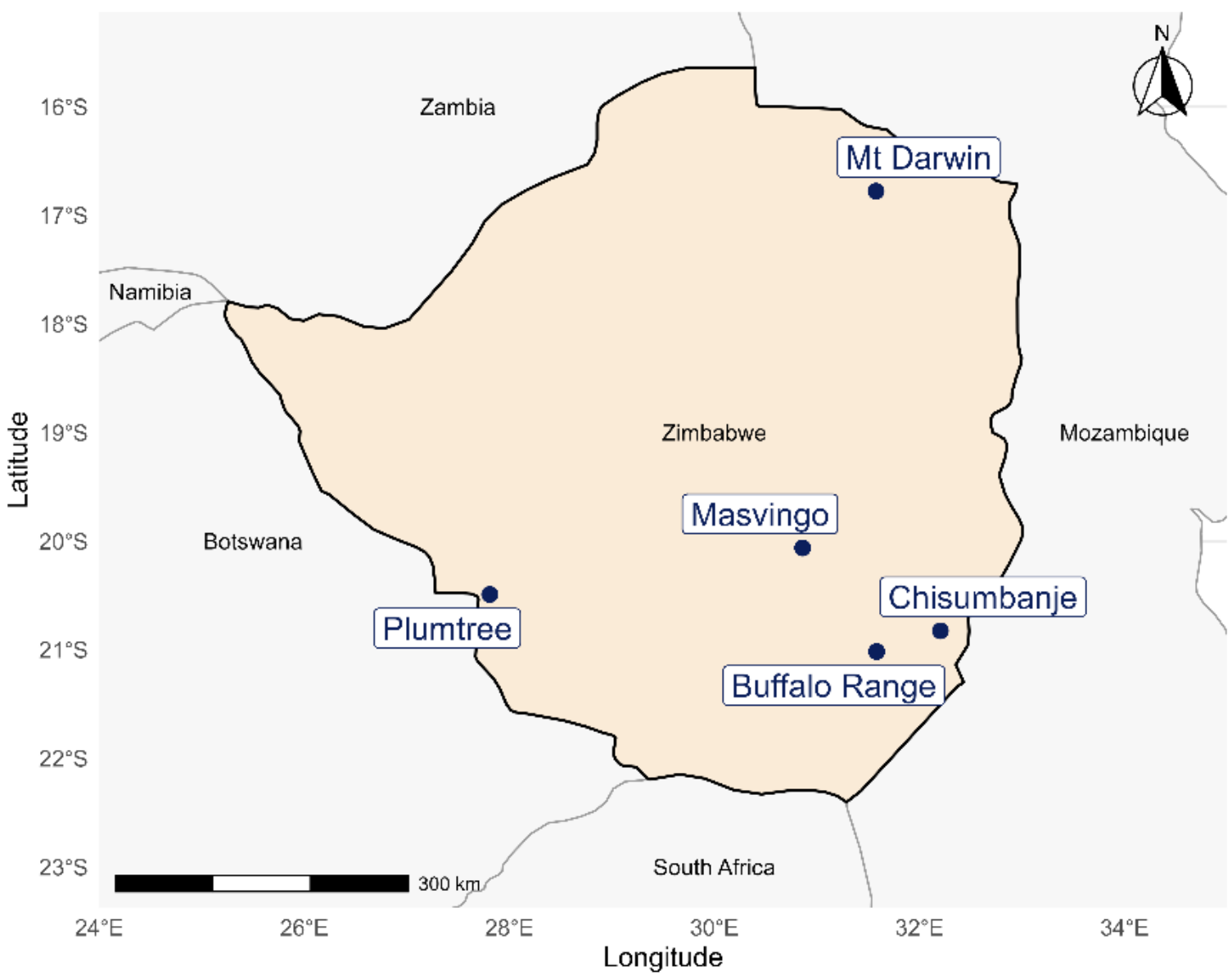

Figure 1: Map of the study area showing location of stations used in Zimbabwe.

# 3 Data

Daily rainfall data from gauge measurements were obtained from the Meteorological Services Department of Zimbabwe for the period 1979 to 2023 for the five stations detailed in Table 1. Prior to analysis, the data were quality controlled using statistical tests adapted from World Meteorological Organization guidelines (World Meteorological Organization, 2021). Range, flat line, and maximum consecutive rain day tests were carried out. A false zeros test was also used to identify entire months during the most rainy period (November to March) where all recorded values are zero, which could indicate missing values incorrectly entered as zeros. These periods were replaced with missing values unless there was strong evidence that the zeros were genuine based on drought information from external sources (Frischen, Meza, Rupp, Wietler, & Hagenlocher, 2020; Meteorological Services Department of Zimbabwe, 2025). The criteria and action applied for each quality control test are provided in Table S2 of SI file 2.

Overall, very few values failed the quality control tests, and the majority of missing values in the records were due to missing records in the original data source.

| Station | Latitude | Longitude | Complete Days (%) |
|---|---|---|---|
| Buffalo Range | -21.02 | 31.58 | 99.4 |
| Chisumbanje | -20.82 | 32.20 | 95.5 |
| Masvingo | -20.06 | 30.86 | 99.8 |
| Mt Darwin | -16.78 | 31.58 | 99.6 |
| Plumtree | -20.48 | 27.80 | 99.8 |

Table 1: Name, location, and percentage of days with a non-missing rainfall value (after quality control) for the five stations in Zimbabwe used in this study.

Precipitation data from the ECMWF Agrometeorological Reanalysis dataset (AgERA5) version 2.0 were used as the input data for the bias correction methods. AgERA5 is derived from the ERA5 reanalysis dataset and provides daily meteorological indicators at a 0.1° spatial resolution globally (Boogaard, et al., 2020). The AgERA5 dataset was designed specifically for use as input to agriculture and agro-ecological models (Boogaard, et al., 2020), hence it is well suited for this study. The AgERA5 precipitation data for the grid pixel closest to each station location (nearest neighbour) was extracted for the same period as the station records. Some differences between the reanalysis and station data may reflect the spatial mismatch between pixel averages and point observations, rather than bias in the data, although the relatively high spatial resolution of AgERA5 (0.1°) may lessen this effect.

# 4 Methodology

## 4.1 Bias Correction Methods

Two widely used bias correction methods for daily rainfall are applied in this study: local intensity scaling (LOCI) and quantile mapping (QM). These methods correct rainfall frequency and amounts separately, making them particularly appealing for many applications. Two new bias correction methods are then introduced, Markov chain LOCI (MC LOCI) and Markov chain QM (MC QM), which integrate Markov chains into the conventional methods, with the aim of improving the temporal structure of daily rainfall values.

Section 4.1.1 describes the LOCI and QM methods, Section 4.1.2 describes MC LOCI and MC QM, and Section 4.1.3 provides a detailed calibration algorithm for the Markov chain methods.

### 4.1.1 Conventional Methods: LOCI and QM

**Local Intensity Scaling (LOCI)**
LOCI is a mean based bias correction method that corrects both rainfall frequency and intensity (Schmidli, Frei, & Vidale, 2006). The LOCI method is a two-step process:

*(1) Adjustment of rainfall occurrence*

The first step ensures that the long term rain day frequency of the bias corrected values matches that of the observation data. To do this, thresholds, $T_Y^m$, are calculated for each month or seasonal period, $m$, such that the probability of the bias corrected data exceeding the threshold in that period matches the frequency of observation data exceeding the standard rain day threshold, $T_X = 0.85$ mm (Stern & Cooper, 2011):

$$P(Y_m(t) > T_Y^m) = P(X_m(t) > T_X) \tag{1}$$

where $X_m(t)$ and $Y_m(t)$ denote observation and model rainfall series in period $m$, respectively.

*(2) Adjustment of rainfall values*

Multiplicative scaling factors $s_m$ are calculated to align the mean intensities of the two datasets:

$$s_m = \frac{E[\,X_m(t) - T_X \mid X_m(t) > T_X\,]}{E[\,Y_m(t) - T_Y^m \mid Y_m(t) > T_Y^m\,]} \tag{2}$$

The LOCI bias corrected series, $Y_m'(t)$ is defined as follows:

$$Y_m'(t) = \begin{cases} 0, & \text{if } Y_m(t) < T_Y^m \\ T_X + s_m(Y_m(t) - T_Y^m), & \text{otherwise} \end{cases} \tag{3}$$

**Quantile Mapping (QM)**

QM is a distribution-based bias correction method that consists of a two-step process. The first step of QM is identical to the first step of LOCI, where $T_Y^m$ values are calculated. The objective of QM is to transform modelled rainfall values using the quantiles of the observation data (Panofsky & Brier, 1968; Ines & Hansen, 2006; Themeßl, Gobiet, & Leuprecht, 2011). QM uses cumulative distribution functions (CDFs) to transform rainfall values. This can be done empirically or parametrically by fitting the data to a suitable distribution, such as a gamma or exponential distribution. We used a Gamma distribution to fit rainfall values, which has been shown to model daily rainfall data well (Kaptué, Hanan, Prihodko, & Ramirez, 2015; McBride, Kruger, & Dyson, 2022).

*Adjustment of rainfall values*

QM uses a quantile-based transformation to correct the full distribution of rainfall values above the calculated threshold. Gamma distributions are fitted separately to observation and model values above their respective thresholds for each month or seasonal period, $m$, as described in Ines & Hansen (2006). The resulting CDFs, $F_{X,m}$ and $F_{Y,m}$ are constructed for the observation and model data respectively.

The bias corrected values, $Y'_m(t)$, are then derived using the inverse CDF (quantile function) of the observations, $F_{X,m}^{-1}$ and the CDF of the model data, $F_{Y,m}$ to map model rainfall values onto the observation CDF:

$$Y'_m(t) = \begin{cases} 0, & \text{if } Y_m(t) < T_Y^m \\ F_{X,m}^{-1}\left(F_{Y,m}\big(Y_m(t)\big)\right), & \text{otherwise} \end{cases} \tag{4}$$

### 4.1.2 Markov chain methods: MC LOCI and MC QM

The Markov chain bias correction methods extend LOCI and QM by incorporating a two-state first order Markov chain in order to improve the temporal structure of rainfall and to provide better representation of wet and dry spells. This is achieved through defining a rain day threshold for each of the two states of the Markov chain and adjusting amounts above the respective threshold dependent on the state of the previous day.

As with LOCI and QM, there are two distinct steps to the methods, adjusting for rainfall occurrence followed by rainfall amounts. The first step is common to both methods.

**(1) Adjustment of rainfall occurrence**

Let $X(t)$ and $Y(t)$ denote the observation and model daily rainfall, respectively. From the observation series, we define a binary rain indicator $W_X(t)$:

$$W_X(t) = \begin{cases} 1, & \text{if } X(t) > T_X \\ 0, & \text{otherwise} \end{cases} \tag{5}$$

where, $T_X = 0.85$ is the standard observation rain day threshold. From this indicator, we compute the unconditional and conditional rain day probabilities, for each month or seasonal period, $m$:

$$p_0^m = P(W_X^m(t) = 1) \tag{6a}$$

$$p_w^m = P(\, W_X^m(t) = 1 \mid W_X^m(t-1) = 1\,) \tag{6b}$$

$$p_d^m = P(\, W_X^m(t) = 1 \mid W_X^m(t-1) = 0\,) \tag{6c}$$

where $W_X^m(t)$ denotes the observation rain day indicator in period $m$.

As in the LOCI and QM methods, the unconditional threshold, $t_0^m$, is calculated such that:

$$P(Y_m(t) > t_0^m) = p_0^m \tag{7}$$

(equal to $T_Y^m$ in LOCI and QM). The conditional probabilities $p_w^m$ and $p_d^m$ describe a first-order Markov chain representation of rainfall occurrence and form the targets for the threshold calibration. We introduce two conditional thresholds, $t_w^m$ and $t_d^m$, used to define rain days in the model data based on the state of the preceding day. The recursive model rain day indicator $W_Y^m(t)$ is defined as:

$$W_Y^m(t) = \begin{cases} 1, & \text{if } W_Y^m(t-1) = 1 \text{ and } Y_m(t) > t_w^m \\ 1, & \text{if } W_Y^m(t-1) = 0 \text{ and } Y_m(t) > t_d^m \\ 0, & \text{otherwise} \end{cases} \tag{8}$$

where $W_Y^m(1) = 1$ if $Y_m(1) > t_0^m$ and 0 otherwise, to initialise the series.

The Markov chain method aims to identify values for $t_w^m$ and $t_d^m$ such that:

$$P(\,Y_m(t) > t_w^m \mid \mathrm{W}_\mathrm{Y}^m(t-1) = 1\,) = p_w^m \tag{9a}$$

$$P(\,Y_m(t) > t_d^m \mid \mathrm{W}_\mathrm{Y}^m(t-1) = 0\,) = p_d^m \tag{9b}$$

i.e. the desired thresholds are defined such that they can be used to ensure that the probability of rainfall given the state of the previous day (rain or no rain) in the bias corrected data matches the corresponding probabilities in the observation data.

However, since the sequence $W_Y^m(t)$ depends on both the threshold values, and altering either threshold alters the probabilities, we cannot directly calculate $t_w^m$ and $t_d^m$ using the quantile function, as is done for the single threshold in LOCI and QM.

From Eq. 9a-9b:

$$t_w^m = \mathrm{Q}_{1-\mathrm{p}_\mathrm{w}^\mathrm{m}}(\,Y_m(t) \mid W_Y^m(t-1) = 1\,) \tag{10a}$$

$$t_d^m = \mathrm{Q}_{1-\mathrm{p}_\mathrm{d}^\mathrm{m}}(\,Y_m(t) \mid W_Y^m(t-1) = 0\,) \tag{10b}$$

where $\mathrm{Q}_\mathrm{k}$ is the quantile function for the probability $k$. Hence, we can consider Eq. 10a-10b as a fixed point equation (Burden & Faires, 2011) where $\theta = (t_w^m, t_d^m)$ is a parameter to be estimated and

$$H(\theta) = \begin{pmatrix} \mathrm{Q}_{1-\mathrm{p}_\mathrm{w}^\mathrm{m}}(\,Y_m(t) \mid W_Y^m(t-1;\theta) = 1\,) \\ \mathrm{Q}_{1-\mathrm{p}_\mathrm{d}^\mathrm{m}}(\,Y_m(t) \mid W_Y^m(t-1;\theta) = 0\,) \end{pmatrix} \tag{11}$$

where we consider $W_Y^m$ as a function of $\theta$ and we are seeking a solution to:

$$\theta = H(\theta) \tag{12}$$

Hence, we can use an iterative approach to find approximate threshold values, analogous to fixed point iteration, widely used in numerical analysis (Burden & Faires, 2011; Huang & Ma, 2014; Puneet, Higinio, Ramandeep, & Vinay, 2023).

A natural choice for initialising is to set: $t_w^m = t_d^m = t_0^m$, since we expect the conditional thresholds to be close to the unconditional threshold.

Once initialised, the following steps are performed iteratively:

- $W_Y^m(t)$ is calculated for the series, as in Eq. 8, based on the current thresholds $(t_w^m, t_d^m)$.
- The exceedance probabilities $\widehat{p_w^m}$ and $\widehat{p_d^m}$ are recomputed, as in Eq. 9a-9b, based on $W_Y^m(t)$ and $(t_w^m, t_d^m)$.
- $t_w^m$ and $t_d^m$ are updated using the quantile function, as in Eq. 10a-10b, towards the conditional quantiles of $Y_m(t)$, with the aim of converging towards $p_w^m$ and $p_d^m$.

The steps are described in detail in 4.1.3.

Since the exceedance probabilities in Eq. 9a-9b are approximately monotonic as functions of the thresholds i.e. increasing a threshold will reduce the number of rain days and hence decrease the exceedance probability, the iterative updates generally will move the solution in the correct direction towards the target probabilities. We also apply a damping factor ($\lambda = 0.4$) to the updates (see 4.1.3) to stabilise the iteration and reduce overshooting and oscillations. The value $\lambda = 0.4$ was selected through preliminary testing, which indicated that it provided stable convergence without requiring an excessive number of iterations.

The iteration terminates when the conditional probabilities are within a defined tolerance ($\epsilon = 0.01$ in 4.1.3) of the target probabilities, or when successive updates become negligible.

Note that because of Markov chain stationarity equation (Ross, 2014):

$$p_0^m = p_w^m p_0^m + p_d^m(1 - p_0^m) \tag{13}$$

we can express $p_0^m$ in terms of $p_w^m$ and $p_d^m$ (Wilks & Wilby, 1999) (assuming the denominator is not equal to zero):

$$p_0^m = \frac{p_d^m}{1 - p_w^m + p_d^m} \tag{14}$$

Hence, if the conditional probabilities $p_w^m$ and $p_d^m$ are matched, then the unconditional probability $p_0^m$ is also matched, and so the overall rain day frequency is maintained by the Markov chain method, matching the observation rain day frequency.

**(2) Adjustment of rainfall amounts**

Following the estimation of the thresholds, rainfall amounts adjustments are defined, conditional on the state of the previous day, using similar definitions to the conventional methods.

*Markov chain local intensity scaling (MC LOCI)*

Conditional scaling factors $s_w^m$ and $s_d^m$ are calculated for each month, as defined in 4.1.3, similar to the LOCI scale factor, but where the rain days of the model and observation data are partitioned into two subsets: rain days following a rain day ($s_w^m$) and rain days following a dry day ($s_d^m$). $s_m$, the standard LOCI scale factor, is also computed for use when the value on the previous day is unavailable. By defining two scale factors, the Markov chain method enables correction of mean rainfall intensity conditional on the state of the previous day. This reflects the commonly observed differences in conditional mean rainfall intensity (Stern & Coe, 1984; Torgbor, Stern, Nkansah, & Stern, 2018), and therefore allows for a more realistic representation of rainfall within wet spells and at the transition between a dry and wet period.

The MC LOCI bias corrected series, $Y_m'(t)$ is defined recursively as follows:

$$Y'_m(1) = \begin{cases} 0, & \text{if } Y_m(1) < T_Y^m \\ T_X + s_m(Y_m(1) - T_Y^m), & \text{otherwise} \end{cases} \tag{15}$$

For subsequent days, if $Y'(t-1) > T_X$:

$$Y'_m(t) = \begin{cases} 0, & \text{if } Y_m(t) \leq t_w^m \\ T_X + s_w^m(Y_m(t) - t_w^m), & \text{if } Y_m(t) > t_w^m \end{cases} \tag{16}$$

If $Y'(t-1) \leq T_X$:

$$Y'_m(t) = \begin{cases} 0, & \text{if } Y_m(t) \leq t_d^m \\ T_X + s_d^m(Y_m(t) - t_d^m), & \text{if } Y_m(t) > t_d^m \end{cases} \tag{17}$$

If $Y'_m(t)$ is missing, $Y'_m(t)$ is calculated using the standard LOCI adjustment (Eq. 3).

Note that the final value from the end of one month or seasonal period can be used as the initial $Y'_m(t-1)$ value for the next month, so the series does not need to be initialised at the start of every month-year period. For example, after calculating $Y'_m(t)$ for January, the value on 31 January can be used as $Y'_m(t-1)$ for 1 February. The exception is for the first month to be initialised. We chose to initialise the series at the start of the dry season to minimise the effect of initialisation on rainy days.

*Markov chain quantile mapping (MC QM)*

As in MC LOCI, the rain days of the model and observation data are partitioned into two subsets: rain days following a rain day, and rain days following a dry day. Gamma distributions are fitted separately to each partition, for the model and observation data, above their respective thresholds. A Gamma distribution is also fitted to all rain days. The CDFs ($F$) and inverse CDFs ($F^{-1}$) are then constructed, as described in 4.1.3. Analogous to MC LOCI, separate rainfall distributions conditional on the state of the previous day allow for differences in rainfall intensities conditional on the previous day states.

The MC QM bias corrected series, $Y'_m(t)$ is defined recursively as follows:

$$Y'_m(1) = \begin{cases} 0, & \text{if } Y_m(1) \leq t_0^m \\ F_X^{-1}\left(F_Y(Y_m(1) - t_0^m)\right) + T_X, & \text{if } Y_m(1) > t_0^m \end{cases} \tag{18}$$

For subsequent days, if $Y'(t-1) > T_X$:

$$Y'_m(t) = \begin{cases} 0, & \text{if } Y_m(t) \leq t_w^m \\ F_{X,w}^{-1}\left(F_{Y,w}(Y_m(t) - t_w^m)\right) + T_X, & \text{if } Y_m(t) > t_w^m \end{cases} \tag{19}$$

If $Y'(t-1) \leq T_X$:

$$Y'_m(t) = \begin{cases} 0, & \text{if } Y_m(t) \leq t_d^m \\ F_{X,d}^{-1}\left(F_{Y,d}(Y_m(t) - t_d^m)\right) + T_X, & \text{if } Y_m(t) > t_d^m \end{cases} \tag{20}$$

where $F$ and $F^{-1}$ functions are as defined in 4.1.3.

If $Y'_m(t-1)$ is missing, $Y'_m(t)$ is calculated using the standard QM adjustment (Eq. 4). This ensures that the correction remains consistent with the unconditional rainfall distribution and avoids assuming a previous wet or dry state.

The final value from the end of one month can be used as the initial $Y'_m(t-1)$ value for the next month to avoid initialisation at the start of every month-year period. We chose to initialise the series at the start of the dry season.

Note that, as with QM, rainfall values can be fitted using the empirical distribution or other appropriate parametric distributions, such as a double exponential distribution.

### 4.1.3 Markov chain bias correction algorithm

The calibration algorithm for the Markov chain bias correction methods described in 4.1.2 is presented in detail in Algorithm 1. This algorithm was developed based on steps of the conventional bias correction methods (4.1.1) and using principles of fixed point equation iteration (Burden & Faires, 2011). The algorithm was implemented in the R statistical language version 4.4.2 (R Core Team, 2021). Full pseudocode for the algorithm is provided in Algorithm S1.

The calibration algorithm was run with eight seasonal calibration periods: dry season (May-September) was considered as a single period, with the remaining seven months from October to April each being an individual period. Recursive iteration was initialised at the beginning of the dry season period (May) to minimise the effect of initialisation on rain days. The rain day threshold used to define wet and dry days in the observation data was $T_x = 0.85$ mm. The iterative threshold calibration used a damping factor of $\lambda = 0.4$, a convergence tolerance of $\epsilon = 0.01$ and a maximum of $N_{max} = 20$ iterations. If convergence was not achieved within $N_{max}$ iterations, the non-convergence was flagged and the final threshold estimates were retained.

Where the previous day's state was missing/unavailable, including for the first day of each validation block, the calculated unconditional rain day threshold ($t_0^m$) was applied to the current day to determine the rain day state. Tied values resulting from quantile calculations were handled using the default linear interpolation method implemented in the R statistical language (type = 7, Hyndman & Fan (1996)). Rainfall values classified as dry days retained a corrected rainfall amount of zero and only non-zero values above the calibrated thresholds were used to calculate scale factors and to fit Gamma distributions.

**Algorithm 1: Markov chain bias correction algorithm**

**Inputs**

- Observation daily series: $X(t)$
- Model daily series: $Y(t)$
- Month/seasonal index: $m = 1, \ldots, M$
- Observation rain day indicator: $W_X(t) = 1$ if $X(t) > T_x$ and 0 otherwise
- Previous day observation indicator: $W_X(t-1)$, computed and stored on the entire series to preserve temporal positioning.
- Convergence tolerance: $\epsilon$
- Max iterations: $N_{max}$
- Damping factor: $\lambda$

**For each month or seasonal period, $m$:**

**Step 1: Extract Seasonal Data**

Subset $X(t)$, $W_X(t)$ and $Y(t)$ for month or seasonal period, $m$ to obtain $X_m(t)$, $W_X^m(t)$ and $Y_m(t)$.

Retain year-block structure so that wet–dry state depends only on the previous day within each sequence.

**Step 2: Estimate Target Occurrence Probabilities from Observations**

From observations $X_m(t)$ compute:

the unconditional rain day probability:

$$p_O^m = P(W_X^m(t) = 1)$$

the conditional rain day probabilities:

$$p_w^m = P(\, W_X^m(t) = 1 \mid W_X^m(t-1) = 1\,)$$

$$p_d^m = P(\, W_X^m(t) = 1 \mid W_X^m(t-1) = 0\,)$$

**Step 3: Initialise Thresholds**

Compute unconditional threshold $t_0^m$ from $Y_m(t)$ such that: $P(Y_m(t) > t_0^m) = p_0^m$

Initialise conditional thresholds: $t_w^m \leftarrow t_0^m$ and $t_d^m \leftarrow t_0^m$

**Step 4: Iterative Threshold Calibration**

Repeat until convergence or $N_{max}$ iterations reached:

*4.1 Generate rain day sequence*

For each year-season block:

Initialise first day:

$W_Y^m(1) \;=\; 1$ if $Y_m(1) > t_0^m$ otherwise 0

For subsequent days:

If $W_Y^m(t\;-1) \;= 1$, then:

$W_Y^m(t) \;= 1$ if $Y_m(t) \; > t_w^m$, otherwise 0

If $W_Y^m(t\;-1) \;= 0$, then:

$W_Y^m(t) \;= 1$ if $Y_m(t) \; > t_d^m$, otherwise 0

If $W_Y^m(t\;-1)$ is missing, then:

$W_Y^m(t) \;=\; 1$ if $Y_m(t) > t_0^m$ otherwise 0

*4.2 Estimate conditional probabilities from new rain day sequence:*

$$\widehat{p_w^m} = P(\, W_Y^m(t) = 1 \mid W_Y^m(t-1) = 1\,)$$

$$\widehat{p_d^m} = P(\, W_Y^m(t) = 1 \mid W_Y^m(t-1) = 0\,)$$

*4.3 Check Convergence:*

Stop if $\left|\widehat{p_w^m} - p_w^m\right| < \epsilon$ and $\left|\widehat{p_d^m} - p_d^m\right| < \epsilon$ or if $\widehat{p_w^m}$ and $\widehat{p_d^m}$ updates have changed by less than $\frac{\epsilon}{2}$.

*4.4 Update Thresholds:*

Update conditional thresholds using quantile function and damping factor $\lambda$ :

$$t_w^{m(new)} = \mathrm{Q}_{1-\mathrm{p}_\mathrm{w}^\mathrm{m}}(Y_m(t) \mid W_Y^m(t-1) = 1)$$

$$t_d^{m(new)} = \mathrm{Q}_{1-\mathrm{p}_\mathrm{d}^\mathrm{m}}(Y_m(t) \mid W_Y^m(t-1) = 0)$$

$$t_w^m \leftarrow (1-\lambda)t_w^m + \lambda t_w^{m(new)}, \quad t_d^m \leftarrow (1-\lambda)t_d^m + \lambda t_d^{m(new)}$$

where $\mathrm{Q}_\mathrm{k}$ is the quantile function for the probability $k$.

**Step 5: Rainfall amounts correction**

MC LOCI: Compute unconditional and conditional scaling factors:

$$s_w^m = \frac{E[\,X_m(t) - T_x \;\mid\; X_m(t) \;>\; T_X \cap W_X^m(t-1) \;=1\,]}{E[\,Y_m(t) - t_w^m \;\mid\; Y_m(t) \;>\; t_w^m \cap W_Y^m(t-1) = 1\,]}$$

$$s_d^m = \frac{E[\,X_m(t) - T_x \;\mid\; X_m(t) \;> T_X \cap W_X^m(t-1) \;=0\,]}{E[\,Y_m(t) - t_d^m \;\mid\; Y_m(t) \;>\; t_d^m \cap W_Y^m(t-1) = 0\,]}$$

$$s_m = \frac{E[\,X_m(t) - T_x \mid X_m(t) > T_X\,]}{E[\,Y_m(t) - t_0^m \mid Y_m(t) > t_0^m\,]}$$

or

MC QM: Fit separate Gamma distributions to the rain days after a rain day and rain days after a dry day subsets and for all wet days. Generate the cumulative distribution functions (CDFs), $F$, and for $X$, the inverse CDFs, $F^{-1}$:

For rain days after a rain day:

$$F_{X,w}^m(x) = P(\,X_m(t) \le x \mid X_m(t) > T_X \cap W_X^m(t-1) = 1\,)$$

$$F_{Y,w}^m(x) = P(\,Y_m(t) \le x \mid Y_m(t) > t_w^m \cap W_Y^m(t-1) = 1\,)$$

For rain days after a dry day:

$$F_{X,d}^m(x) = P(\,X_m(t) \le x \mid X_m(t) > T_X \cap W_X^m(t-1) = 0\,)$$

$$F_{Y,d}^m(x) = P(\,Y_m(t) \le x \mid Y_m(t) > t_d^m \cap W_Y^m(t-1) = 0\,)$$

For all rain days:

$$F_X^m(x) = P(\,X_m(t) \le x \mid X_m(t) > T_X\,)$$

$$F_Y^m(x) = P(\,Y_m(t) \le x \mid Y_m(t) > t_0^m\,)$$

**Step 6: Store Results**

For each month or seasonal period, store:

- Calibrated thresholds: $t_0^m, t_w^m, t_d^m$
- Target and achieved probabilities: $p^m, p_w^m, p_d^m$ and $\widehat{p_w^m}, \widehat{p_d^m}$
- For MC LOCI: calibrated scale factors: $s_w^m$, $s_d^m$ and $s_m$
- For MC QM: the inverse CDFs for $X$ and CDFs for $Y$: $\left(F_{X,w}^m\right)^{-1}$, $F_{Y,w}^m$, $\left(F_{X,d}^m\right)^{-1}$, $F_{Y,d}^m$, $(F_X^m)^{-1}$, $F_Y^m$

## 4.2 Evaluation Methodology

The parameters for the four bias correction methods were calculated using a block cross-validation approach. For each location, the AgERA5 and gauge datasets were divided into four blocks (1979-1988, 1989-1998, 1999-2008, 2009-2023). For each of the four blocks, the remaining three blocks were used as the calibration period. For all methods the bias correction parameters (thresholds, probabilities, scale factors and fitted distributions) were estimated exclusively from the calibration period, and then applied to the held-out validation block. This produced four independent validation datasets, which were combined into a single validation time series for the period 1979-2023. The block cross-validation approach reduces any effects of non-stationarity over time by allowing parameters to be estimated using data from throughout the record rather than from a single period.

The time series for the bias corrected data for each location were compared with the unadjusted AgERA5 data and the gauge data. Evaluation was separated into two components: rainfall occurrence evaluation and rainfall amounts evaluation. Table 2 provides an overview of the statistical methods and summaries used for each component, categorised into four sections: climatology, annual summaries, distributions and detection skill. Metrics were calculated across the combined validation series. Key performance metrics related to spell length distributions and temporal structure were also evaluation across station-validation blocks to assess block-wise variability. The evaluation framework was adapted from Parsons et al. (2026) and allowed for a comprehensive evaluation of rainfall characteristics over a range of temporal scales. The target and achieved transition probabilities were also compared for each validation block to evaluate the stability and robustness of the Markov chain bias correction methods and variability between blocks.

| | Rainfall occurrence | Rainfall amounts |
|---|---|---|
| Climatology (long-term monthly) | Mean number of rain days | Mean total rainfall<br>Mean of mean rainfall per rain day<br>Mean maximum daily rainfall |
| Annual summaries (August-July) | Total number of rain days<br>Longest dry spell (October to March) | Total rainfall<br>Mean rainfall per rain day<br>Maximum daily rainfall |
| Distributions | Distribution of wet and dry spells (October to March)<br>Zero and first-order Markov chain models of rainfall occurrence | Zero and first-order Markov chain models of mean rainfall per rain day |
| Detection skill | Rain vs no rain detection | Rainfall intensity category detection |

Table 2: An overview of the statistical methods and summaries used in the evaluation of the bias correction methods.

The rainfall occurrence evaluation analysed the binary rain day indicator variable, defined using a threshold of 0.85mm for a rain day. 0.85mm is used to reduce potential bias in gauge records due to rounding and is practically equivalent to the standard 1mm threshold (Stern & Cooper, 2011), and was also applied to AgERA5 data for consistency. This threshold has been applied in previous studies of rainfall records in southern Africa (Stern & Cooper, 2011) and other regions (Froidurot & Diedhiou, 2017).

To evaluate the climatology values and annual summaries, four statistical metrics were used, as defined in Table 3. These metrics evaluate systematic error (mean error), accuracy (root mean square error), association (correlation coefficient) and variability (ratio of standard deviations), allowing the bias corrected values to be evaluated across multiple aspects of performance.

| **Metric** | **Formula** | **Range** | **Ideal value** |
|---|---|---|---|
| Mean error | $\frac{1}{N}\sum_{i=1}^{N}(Y_i - X_i)$ | –∞ to ∞ | 0 |
| Root mean square error | $\sqrt{\frac{1}{N}\sum_{i=1}^{N}(Y_i - X_i)^2}$ | 0 to ∞ | 0 |
| Correlation Coefficient | $\frac{\sum_{i=1}^{N}(Y_i - \mu_Y)(X_i - \mu_X)}{\sqrt{\sum_{i=1}^{N}(Y_i - \mu_Y)^2 \sum_{i=1}^{N}\left(X_i - \mu_X\right)^2}}$ | –1 to 1 | 1 |
| Ratio of standard deviations | $\frac{\sigma_Y}{\sigma_X}$ | 0 to ∞ | 1 |

Table 3: Metrics used for evaluation of climatology values and annual summaries. $Y_i$ is the model value at time $i$, $X_i$ is the gauge data value at time $i$, $N$ is the length of the series, $\mu_Y$ and $\mu_X$ are the mean of the model and gauge data respectively and $\sigma_Y$ and $\sigma_X$ are the standard deviations of the model and gauge data respectively.

The lengths of all wet spells and dry spells from October-March were calculated and the distributions from the bias corrected data were compared with the corresponding distributions from the gauge data using the two sample non-parametric Kolmogorov-Smirnov (K-S) test statistic, $D$, defined as the maximum vertical difference between the empirical cumulative distribution functions (ECDFs):

$$D = \sup_x \left| F_X(x) - F_Y(x) \right| \tag{21}$$

where $F_X(x)$ and $F_Y(x)$ are the ECDFs of wet (dry) spell lengths from the gauge data and the reanalysis or bias corrected data, respectively. The test statistic $D$ was used to test a null hypothesis of whether the two samples were from the same distribution.

Markov chain models were used to analyse and compare the seasonal patterns in rainfall occurrence and intensity. The zero-order rainfall occurrence Markov chain model estimates the probability of rainfall as a function of the day of the year using logistic regression with Fourier series terms to represent seasonality. The first-order rainfall occurrence Markov chain model additionally includes the state of the previous day (rain or no rain) as a predictor, allowing for analysis of rainfall onset and persistence. Equivalent models were fitted only on rain days to estimate rainfall amounts using a generalised linear model with a Gamma distribution and Fourier series terms to represent seasonality. These models have been described and used extensively in the analysis of daily rainfall data (Gabriel & Neumann, 1962; Stern & Coe, 1984; Jimoh & Webster, 1996; Torgbor, Stern, Nkansah, & Stern, 2018; da Silva Jale, et al., 2019). Models for gauge, reanalysis, and bias corrected data were fitted separately, restricted to days with available data for all sources.

To quantify the difference between Markov chain models for different data sources, we defined a metric based on the root mean square error (RMSE) of the fitted values throughout the season. For zero-order Markov chain models the metric is defined as:

$$RMSE_{curve}^{(0)} = \sqrt{\frac{1}{366}\sum_{d=1}^{366}(r_d^Y - r_d^X)^2} \quad (22)$$

where $r_d^X$ and $r_d^Y$ represent the fitted values for day $d$ from the zero-order Markov chain model for the gauge data and reanalysis or bias corrected data, respectively. For rainfall occurrence models, $r_d$ values correspond to fitted probability of rainfall, while for rainfall intensity models, $r_d$ values correspond to fitted rainfall amounts.

For first-order Markov chain models, two metric values are defined similarly, one for each fitted curve:

$$RMSE_{curve}^{(i)} = \sqrt{\frac{1}{366}\sum_{d=1}^{366}\left(r_{i,d}^Y - r_{i,d}^X\right)^2} \quad (23)$$

where $i \in \{W, D\}$. The case $i = W$ corresponds to fitted values conditional on the previous day being wet, and $i = D$ corresponds to fitted values conditional on the previous day being dry. The terms $r_{i,d}^X$ and $r_{i,d}^Y$ denote the fitted values for day $d$ from the gauge data and the reanalysis or bias corrected data, respectively.

Rainfall detection skill was evaluated by calculating probability of detection (POD), false alarm ratio (FAR) and Heidke Skill Score (HSS) for the binary rain day indicator, and POD and HSS for the categorical rainfall intensity indicator (Table 4). These metrics allow evaluation of the ability to correctly detect rainfall events (POD) and the frequency of incorrect rain detections (FAR), as well as overall skill (HSS). The five rainfall categories were defined as: dry (x < 0.85mm), light rain (0.85mm <= x < 5mm), moderate rain (5mm <= x < 20mm), heavy rain (20mm <= x < 40mm) and violent rain (x >= 40mm) where x denotes daily rainfall. This categorisation was adapted from WMO intensity classifications (Acharya, Nathan, Wang, Su, & Eizenberg, 2019; Zambrano-Bigiarini, Nauditt, Birkel, Verbist, & Ribbe, 2017).

| **Metric** | **Formula (rain day indicator)** | **Formula (categorical rainfall intensity indicator)** | **Range** | **Ideal value** |
|---|---|---|---|---|
| Probability of Detection (POD) | $\frac{H}{H+M}$ | $POD_i = \frac{n_{ii}}{n_{.i}}$ per category $i$ | 0 to 1 | 1 |
| False Alarm Ratio (FAR) | $\frac{F}{H+F}$ | - | 0 to 1 | 0 |
| Heidke Skill Score | $\frac{2(H \cdot C - F \cdot M)}{(H+M)(M+C)+(H+F)(F+C)}$ | $\frac{\sum_i n_{ii} - \sum_i \frac{n_{i.}n_{.i}}{N}}{N - \sum_i \frac{n_{i.}n_{.i}}{N}}$ | –∞ to 1 | 1 |

Table 4: Detection metrics used for rainfall occurrence and categorical rainfall intensity indicators. $N$ = total number of observations. For rain day indicator: $H$ = number of hits (model = rain & gauge = rain), $M$ = number of misses (model = no rain & gauge = rain), $C$ = number of correct negatives (model = no rain & gauge = no rain), $F$ = number of false alarms (model = rain & gauge = no rain). For categorical rainfall intensity indicator with $k$ categories: $n_{ij}$ = number of cases with model category $i$ and gauge category $j$, $n_{i.} = \sum_j n_{ij}$, $n_{.j} = \sum_i n_{ij}$.

# 5 Results

## 5.1 Calibration of Markov chain probabilities

Figure 2 shows the set of calibrated versus target probability values for the Markov chain bias correction methods, for both the conditional probabilities, $p_w$ and $p_d$ and the overall unconditional probability of rainfall, $p_0$, across all stations, months, and calibration blocks. The $y = x$ reference line is shown to indicate how close the calibrated probabilities are to the target probabilities. Figure 2 shows that the calibrated probabilities were close to the target probabilities across all stations, months, and calibration blocks. This finding was seen for both the conditional probabilities, which the Markov chain bias correction method is designed to match, and for the unconditional probabilities, which is achieved when both conditional probabilities are matched. A clear distinction between $p_w$ and $p_d$ values was also observed (Figure 2), and the corresponding conditional thresholds, $t_w$ and $t_d$ differed by between 0 and 5.55 mm (Table S1 of Supplementary Information (SI) file 1).

The largest deviations between target and calibrated probabilities occurred for $p_w$ values in the dry season and in the neighbouring months of October and April (Table S1). These periods have the lowest number of rain after rain days, resulting in the smallest sample size for calibration and therefore the largest deviations.

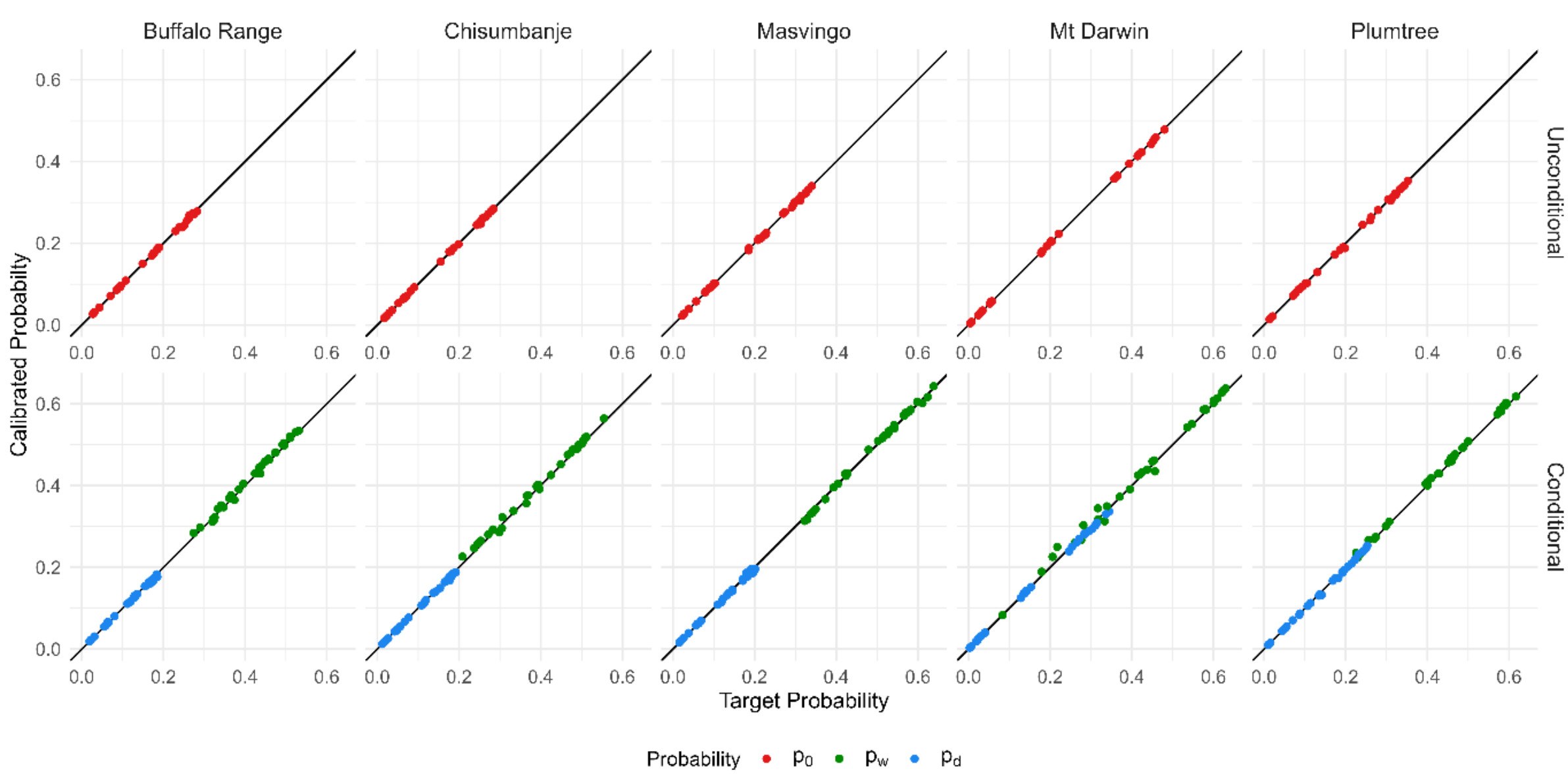

Figure 2: Calibrated versus target probability values of the Markov chain bias correction method for conditional probabilities, $p_w$ and $p_d$ (bottom row), and overall unconditional probability of rainfall, $p_0$ (top row), for the four calibration blocks at each station.

## 5.2 Rainfall occurrence

Since the LOCI and QM methods share the same first step (defining a rain day threshold), their sequences of rain day indicators are identical. Hence, for the rainfall occurrence analysis in this section, they are labelled together as LOCI/QM. The same applies to MC LOCI and MC QM, which are labelled together as MC in this section.

Note that this grouping only applies for rainfall occurrence evaluation and  the methods are distinguished again in the rainfall amounts evaluation.

The majority of results in this section are reported at the station level. To assess the robustness of the bias correction methods across locations and validation periods, the mean and standard deviation of key performance metrics related to spell length distributions and temporal structure were also calculated across the 20 station-validation block combinations and are presented in Table 5.

| Metric | AgERA5 | LOCI/QM | MC |
|---|---|---|---|
| K-S test statistic - wet spells | 0.216 (0.051) | 0.095 (0.040) | 0.058 (0.025) |
| K-S test statistic - dry spells | 0.116 (0.048) | 0.095 (0.034) | 0.061 (0.020) |
| $RMSE_{curve}^{(0)}$ | 0.161 (0.026) | 0.141 (0.032) | 0.141 (0.032) |
| $RMSE_{curve}^{(D)}$ | 0.061 (0.022) | 0.024 (0.010) | 0.013 (0.004) |
| $RMSE_{curve}^{(W)}$ | 0.212 (0.048) | 0.074 (0.040) | 0.051 (0.027) |

Table 5: Mean (standard deviation) values of key rainfall occurrence performance metrics (Kolmogorov-Smirnov (K-S) test statistics for wet and dry spells, Root Mean Square Error (RMSE) values for Markov chain models of rainfall occurrence) across the 20 station–validation block combinations for AgERA5, LOCI/QM, and MC methods.

### 5.2.1 Climatology

AgERA5 captured the seasonal pattern of rain days but consistently overestimated rainfall occurrence at every station in each month (Figure 3). Both LOCI/QM and MC corrected this bias equally well, producing monthly means that closely matched the gauge data.

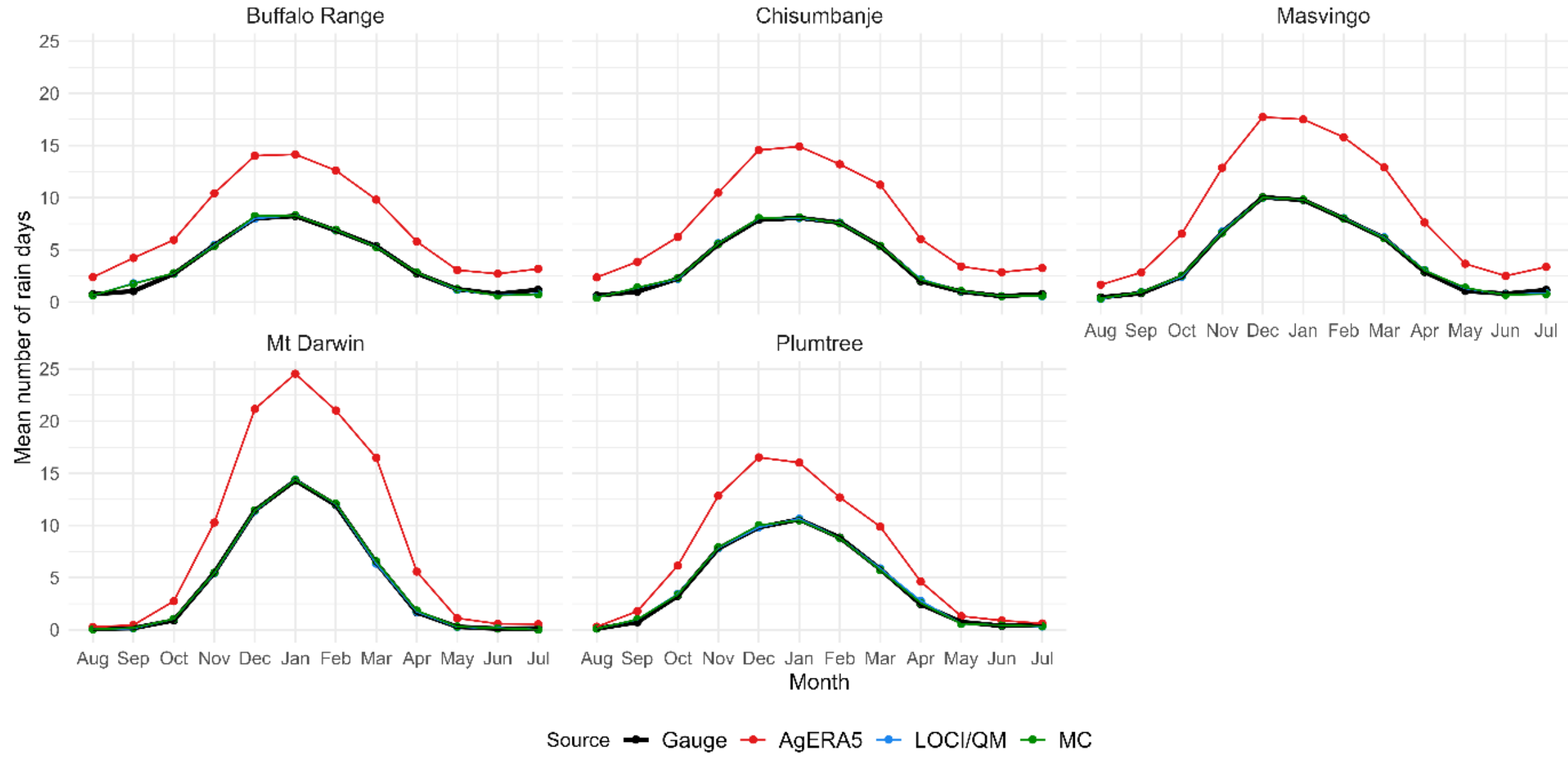

Figure 3: Mean number of rain days in each month from at the five stations according to AgERA5, LOCI/QM, and MC compared with the gauge data.

### 5.2.2 Annual summaries

**Number of rain days**

AgERA5 consistently overestimated the annual number of rain days at every station, with a mean error of 32-54 days. Correlation was moderate to strong (r = 0.79-0.86) at four stations, as shown by the matching peaks and troughs in Figure 4. Plumtree was an exception, where the correlation was 0.19 (Table 6).

LOCI/QM substantially reduced the mean error to between -1.38 and 0.61 days across stations (Table 5). Correlation for LOCI/QM was largely unchanged compared with AgERA5, with similar values or slight decreases at some stations (Table 6). MC further reduced the mean error slightly at four of the five stations. Correlation values for MC were similar to those of LOCI/QM, and Plumtree remained poorly correlated under both methods (r = 0.02-0.04).

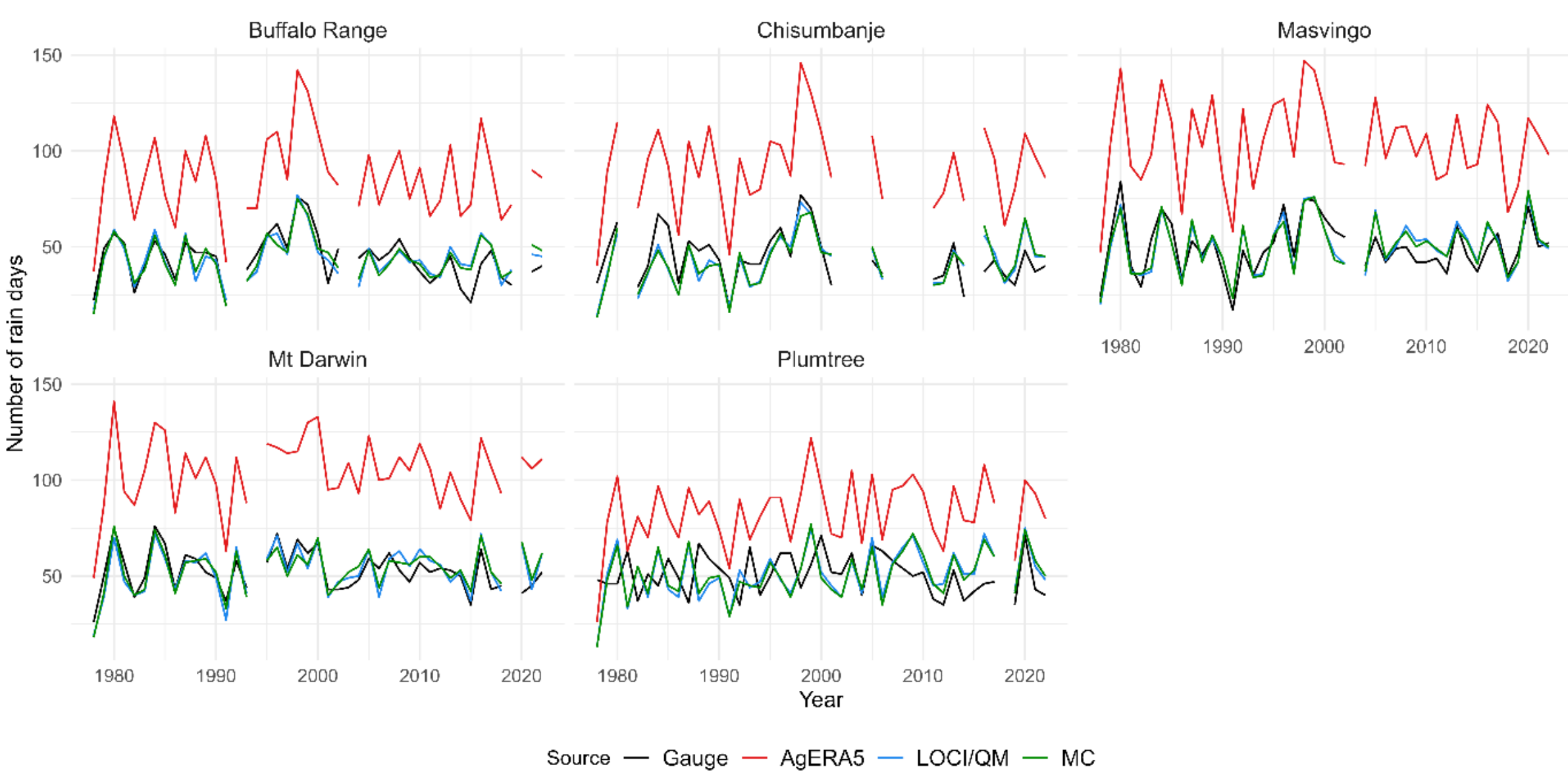

Figure 4: Annual number of rain days at the five stations from Gauge, AgERA5, LOCI/QM, and MC sources.

| Station | Mean Error | | | Correlation | | |
|---|---|---|---|---|---|---|
| | AgERA5 | LOCI/QM | MC | AgERA5 | LOCI/QM | MC |
| Buffalo Range | 42.95 | -0.21 | **-0.05** | **0.84** | 0.80 | 0.82 |
| Chisumbanje | 47.38 | -1.38 | **-1.35** | **0.79** | 0.70 | 0.66 |
| Masvingo | 54.16 | 0.34 | **0.16** | **0.86** | 0.85 | 0.84 |
| Mt Darwin | 51.60 | **0.28** | 0.84 | 0.80 | 0.80 | **0.81** |
| Plumtree | 32.48 | 0.61 | **0.34** | **0.19** | 0.02 | 0.04 |

Table 6: Mean error and correlation values for annual number of rain days by AgERA5, LOCI/QM, and MC in comparison to the gauge data for the five stations. Bold values indicate the data source closest to the metric's ideal value for each station.

**Length of longest dry spell (October to March)**

AgERA5 consistently underestimated the longest dry spell between October and March, with mean errors of -13.39 to -4.84 days (Table 7). Correlation was weak positive to no correlation at all stations. LOCI/QM reduced the systematic bias, but maintained a small overestimation at each station, with mean errors of 1.64-3.68 days. Correlation remained weak, with a small increase at one station. Figure 5 shows that LOCI/QM substantially overestimated the longest dry spell on a small number of occasions. A similar large overestimation occurred on one occasion at Mt Darwin for MC.

MC further reduced the mean error to -0.51 to 1.07 days, slightly outperforming LOCI/QM across all stations. Correlation was similar to LOCI/QM and AgERA5, with a small decrease at Mt Darwin and Chisumbanje and a small increase at Buffalo Range.

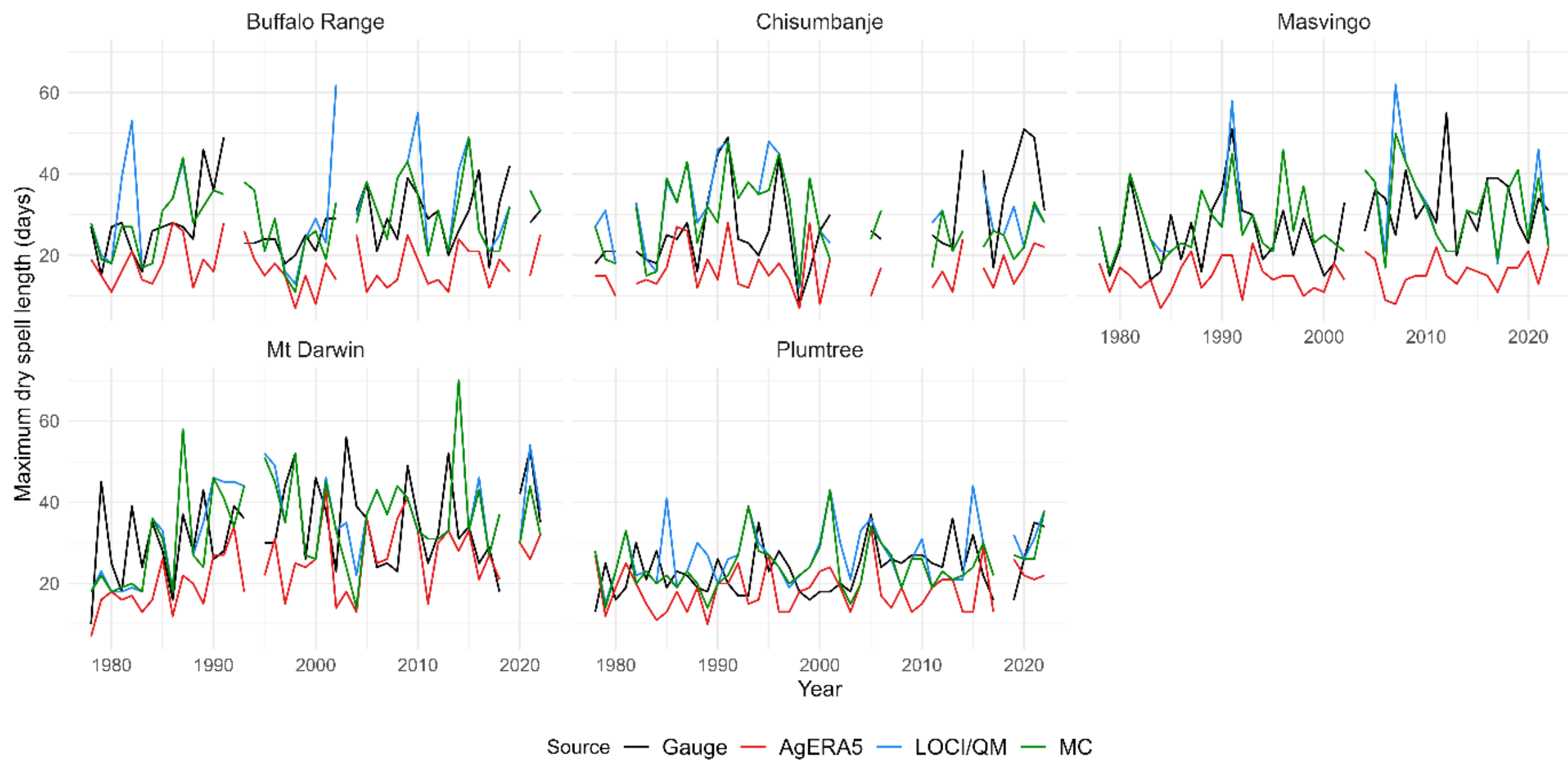

Figure 5: Annual length of longest dry spell at the five stations from Gauge, AgERA5, LOCI/QM, and MC sources.

| Station | Mean Error | | | Correlation | | |
|---|---|---|---|---|---|---|
| | AgERA5 | LOCI/QM | MC | AgERA5 | LOCI/QM | MC |
| Buffalo Range | -10.86 | 3.26 | **0.64** | 0.40 | 0.33 | **0.45** |
| Chisumbanje | -12.08 | 2.05 | **-0.51** | **0.41** | 0.33 | 0.21 |
| Masvingo | -13.39 | 1.64 | **0.75** | 0.19 | **0.36** | 0.35 |
| Mt Darwin | -9.35 | 2.63 | **0.98** | **0.30** | 0.25 | 0.13 |
| Plumtree | -4.84 | 3.68 | **1.07** | 0.02 | **0.05** | **0.05** |

Table 7: Mean error and correlation values for annual length of longest dry spell by AgERA5, LOCI/QM, and MC in comparison to the gauge data for the five stations. Bold values indicate the data source closest to the metric's ideal value for each station.

### 5.2.3 Distribution of wet and dry spells

**Distribution of wet spells (October to March)**

AgERA5 underestimated the frequency of wet spells of varying lengths from October to March (Figure 6). The K-S test statistics were 0.176-0.282, with highly significant P values at all stations (Table 8), indicating significant differences in distribution for AgERA5 compared with the gauge data. LOCI/QM reduced the underestimation, with K-S statistics of 0.055-0.123. However, wet spell lengths remained slightly underestimated at all stations. The K-S tests were highly significant at four of the five stations (the exception was Masvingo), indicating significant differences from the gauge distribution at these four stations.

MC produced lower K-S statistics than AgERA5 and LOCI/QM at every station (0.023-0.050). Wet spells were slightly underestimated but by a substantially smaller amount than for AgERA5 and LOCI/QM. The K-S test statistics were not significant at any station, indicating no statistical evidence of a difference between the MC and gauge wet spell distributions.

Across the 20 station-validation block combinations, the MC method achieved the lowest mean K–S statistic for wet-spell lengths (0.058), compared with 0.095 for LOCI/QM and 0.216 for AgERA5 (Table 5). The standard deviation was also lowest for the MC method (0.025).

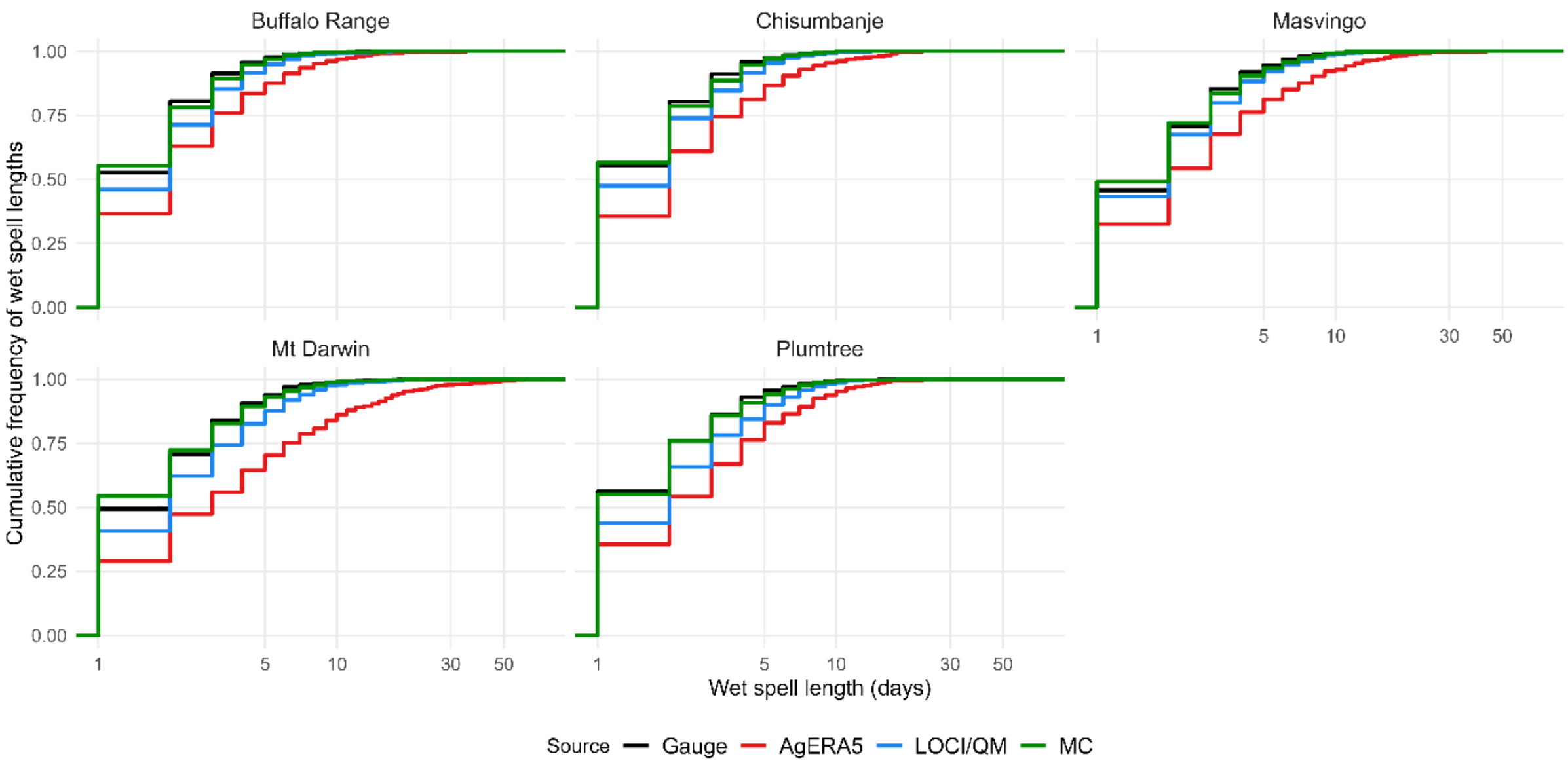

Figure 6: Cumulative frequency graphs of wet spells lengths at the five stations from Gauge, AgERA5, LOCI/QM, and MC sources.

| Station | Source | K-S test statistic | P value |
|---|---|---|---|
| Buffalo Range | AgERA5 | 0.176 | <0.001 |
| | LOCI/QM | 0.093 | 0.002 |
| | **MC** | **0.025** | 0.948 |
| Chisumbanje | AgERA5 | 0.198 | <0.001 |
| | LOCI/QM | 0.079 | 0.016 |
| | **MC** | **0.025** | 0.964 |
| Masvingo | AgERA5 | 0.177 | <0.001 |
| | LOCI/QM | 0.055 | 0.169 |
| | **MC** | **0.033** | 0.735 |
| Mt Darwin | AgERA5 | 0.282 | <0.001 |
| | LOCI/QM | 0.098 | <0.001 |
| | **MC** | **0.050** | 0.156 |
| Plumtree | AgERA5 | 0.219 | <0.001 |
| | LOCI/QM | 0.123 | <0.001 |
| | **MC** | **0.023** | 0.958 |

Table 8: Kolmogorov-Smirnov (K-S) test statistics and P values for the frequency of wet spells from October to March for AgERA5, LOCI/QM, and MC compared to the Gauge data for the five stations. Bold values indicate the data source with the lowest K-S test statistic value for each station.

**Distribution of dry spells (October to March)**

AgERA5 overestimated the frequency of dry spells of varying lengths from October to March at all stations (Figure 7), with the difference increased for dry spells exceeding five days. The overestimation was small at Mt Darwin and Plumtree (K-S statistics 0.043 and 0.068, respectively), and larger at the remaining three stations (0.121-0.166, Table 9). P values were highly significant at three stations and significant at Plumtree, indicating that the AgERA5 dry spell distribution was significantly different from that of the gauge data. At Mt Darwin, the difference was not significant.

LOCI/QM underestimated the cumulative frequency of dry spells, indicating overcorrection of the AgERA5 bias (Figure 7). The K-S statistic was reduced compared with AgERA5 at Buffalo Range, Chisumbanje, and Masvingo, but was increased at Mt Darwin and Plumtree. At Masvingo, the distribution of dry spells was close to the gauge data, and the P value confirmed no statistical evidence of a difference in distribution ($p = 0.338$, Table 9). At the other four stations significant differences were observed, indicating that the LOCI/QM distribution of dry spells was significantly different from the gauge distribution.

The distribution dry spells of MC most closely matched the gauge distribution at four of the five stations (Figure 7). Except at Mt Darwin, the K-S statistics were lower than AgERA5 and LOCI/QM and produced non-significant P values at these four stations. This means that we cannot reject the null hypothesis that the distribution of dry spells from the gauge and MC data have the same distribution at these four stations. At Mt Darwin MC outperformed LOCI/QM but did not achieve a non-significant result (K-S=0.067, $p=0.017$), and AgERA5 showed the closest match of the three.

The station-validation block summary (Table 5) showed a similar overall pattern, with the MC method having the lowest mean K-S statistic (0.061) and lowest standard deviation among the evaluated methods (Table 5).

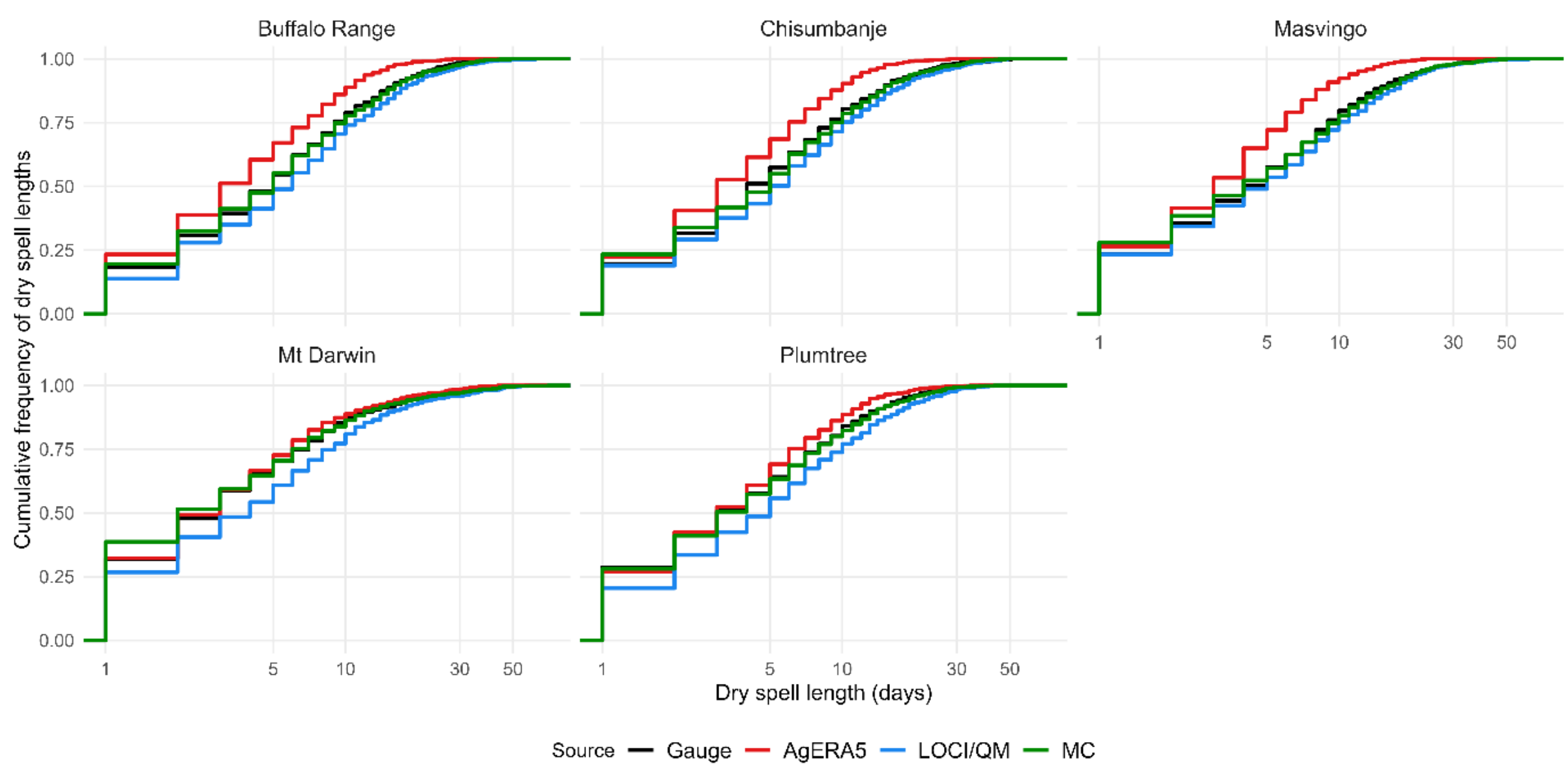

Figure 7: Cumulative frequency graphs of dry spells lengths at the five stations from Gauge, AgERA5, LOCI/QM, and MC sources.

| Station | Source | K-S test statistic | p value |
|---|---|---|---|
| Buffalo Range | AgERA5 | 0.124 | <0.001 |
| | LOCI/QM | 0.069 | 0.037 |
| | MC | **0.018** | 0.998 |
| Chisumbanje | AgERA5 | 0.121 | <0.001 |
| | LOCI/QM | 0.079 | 0.013 |
| | MC | **0.037** | 0.585 |
| Masvingo | AgERA5 | 0.166 | <0.001 |
| | LOCI/QM | 0.046 | 0.338 |
| | MC | **0.045** | 0.333 |
| Mt Darwin | AgERA5 | **0.043** | 0.375 |
| | LOCI/QM | 0.109 | <0.001 |
| | MC | 0.067 | 0.017 |
| Plumtree | AgERA5 | 0.068 | 0.017 |
| | LOCI/QM | 0.092 | <0.001 |
| | MC | **0.018** | 0.996 |

Table 9: Kolmogorov-Smirnov (K-S) test statistics and P values for the frequency of dry spells from October to March for AgERA5, LOCI/QM, and MC compared to the Gauge data for the five stations. Bold values indicate the data source with the lowest K-S test statistic value for each station.

### 5.2.4 Seasonal distribution

**Zero-order Markov chain models of rainfall occurrence**

The zero-order Markov chain model graphs in Figure 8 show the estimated probability of rainfall for each day of the year from August to July. AgERA5 reproduced the seasonal pattern of rain day probability but overestimated rainfall occurrence at all stations. During the peak rainy season (November to February), probabilities AgERA5 were frequently double those of the gauge data.

LOCI/QM corrected the bias and closely matched the gauge curves. MC gave an almost identical correction and performed equally well (Figure 8). The $RMSE_{curve}^{(0)}$ for AgERA5 ranged from 0.114 to 0.197, whereas for LOCI/QM and MC, values were similar < 0.01. (Table 10). The station-validation block mean and standard deviation of $RMSE_{curve}^{(0)}$ were also the same for LOCI/QM and MC (Table 5).

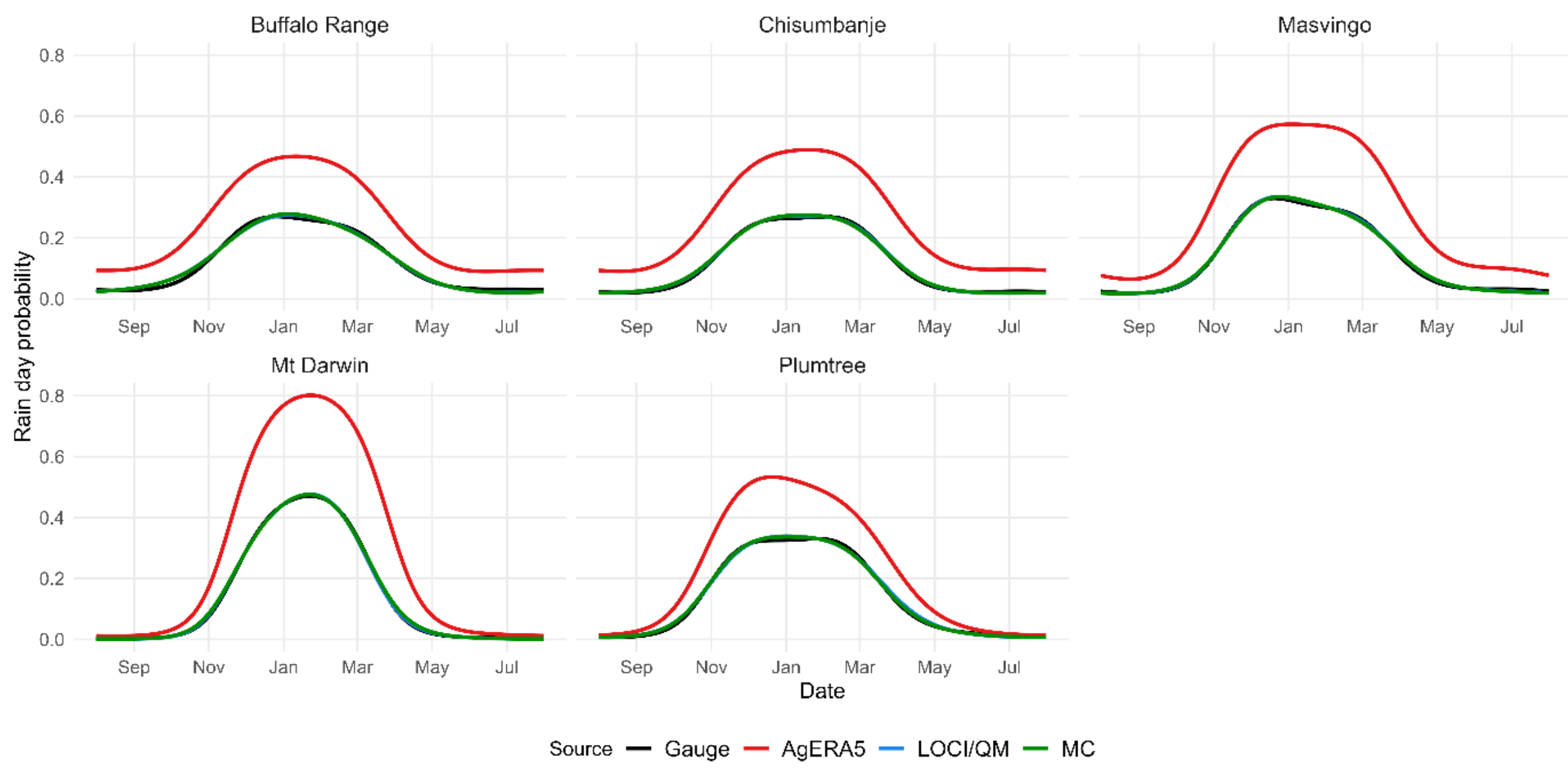

Figure 8: Zero-order Markov chain model curves showing the estimated probability of rainfall for each day of the year, from August to July at the five stations, for Gauge, AgERA5, LOCI/QM, and MC sources.

| Station | AgERA5 | LOCI/QM | MC |
|---|---|---|---|
| Buffalo Range | 0.131 | 0.008 | 0.008 |
| Chisumbanje | 0.148 | 0.005 | 0.005 |
| Masvingo | 0.171 | 0.005 | 0.006 |
| Mt Darwin | 0.197 | 0.003 | 0.004 |
| Plumtree | 0.114 | 0.007 | 0.005 |

Table 10: Root Mean Square Error ($RMSE_{curve}^{(0)}$) values for probability of rainfall for AgERA5, LOCI/QM, and MC compared to the gauge data for the five stations.

**First-order Markov chain models of rainfall occurrence**

The first-order Markov chain model graphs in Figure 9 show the estimated probability of rainfall conditional on the state of the previous day (rain or no rain) for each day of the year from August to July. AgERA5 reproduced a similar seasonal pattern to the gauge data for both states. Similar to the gauge data, AgERA5 rain day probabilities were typically 2-3 times greater when the previous day was wet compared to when it was dry. However, AgERA5 overestimated probabilities for both states, with larger overestimation when the previous day was wet.

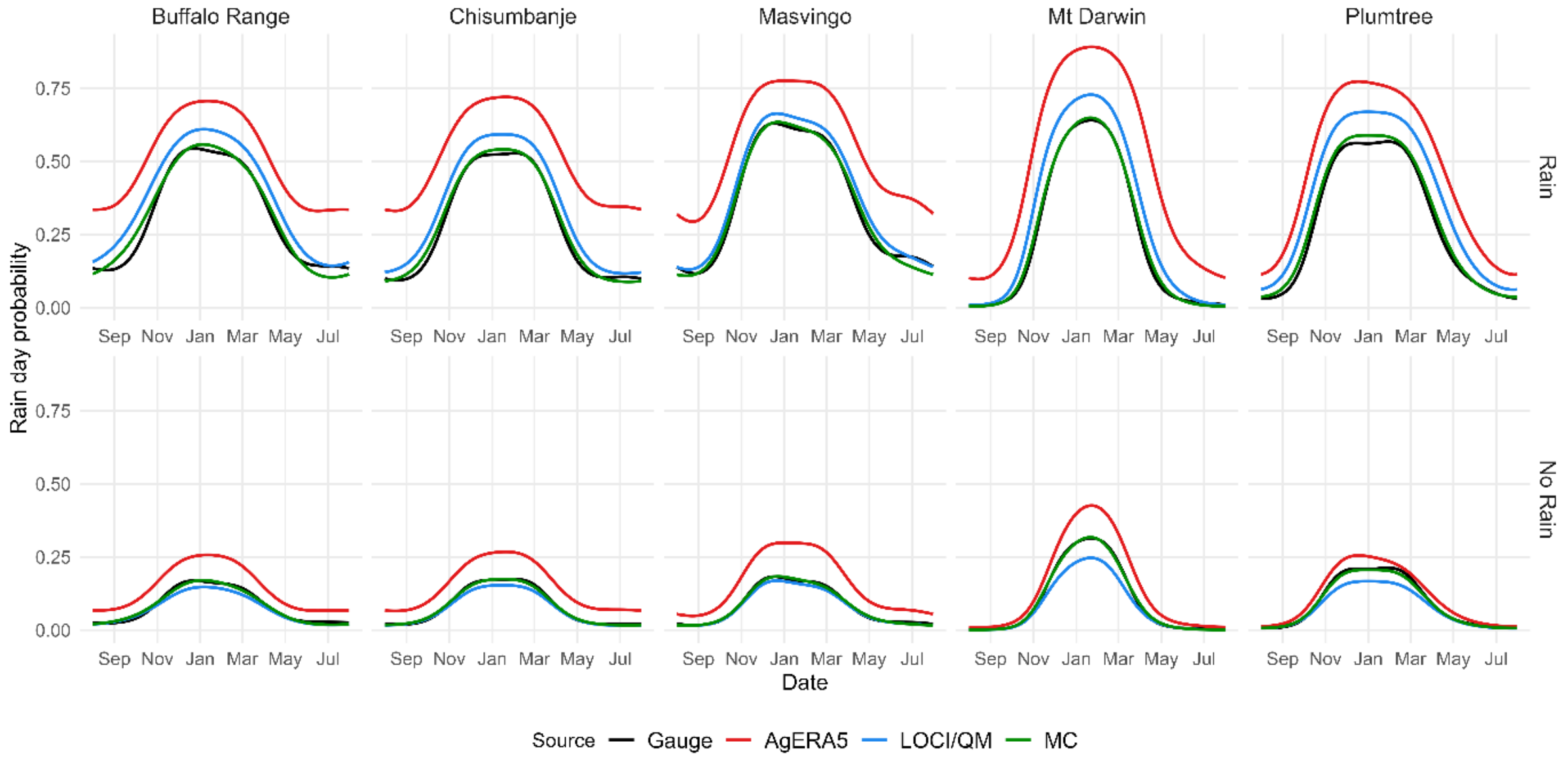

Figure 9: First-order Markov chain model curves showing the estimated probability of rainfall with separate curves conditional on the state of the previous day (rain or no rain), for each day of the year, from August to July at the five stations, for Gauge, AgERA5, LOCI/QM, and MC sources.

LOCI/QM reduced the bias while maintaining a similar seasonal pattern to the gauge data (Figure 9). When the previous day was wet, LOCI/QM slightly overestimated the probability of a rain day, with a reduced $RMSE_{curve}^{(W)}$ values of 0.04-0.1 (Figure 10). When the previous day was dry, LOCI/QM slightly underestimated rain day probability, with $RMSE_{curve}^{(D)}$ values of 0.01-0.02. The largest underestimation was at Mt Darwin and Plumtree, consistent with the overcorrection of the distribution of dry spells.

MC closely matched the gauge curves for both states and outperformed LOCI/QM consistently across all stations. $RMSE_{curve}^{(W)}$ values were below 0.03 and $RMSE_{curve}^{(D)}$ were below 0.01 at all stations (Figure 10), an improvement over LOCI/QM. The values from Figure 10 are shown in Table S3. This is consistent with the station-validation block summary, where the mean $RMSE_{curve}^{(W)}$ and $RMSE_{curve}^{(D)}$ values were reduced from 0.074 to 0.051 and from 0.024 to 0.013, respectively, for MC relative to LOCI/QM (Table 5). The standard deviation values were also lower for the MC method.

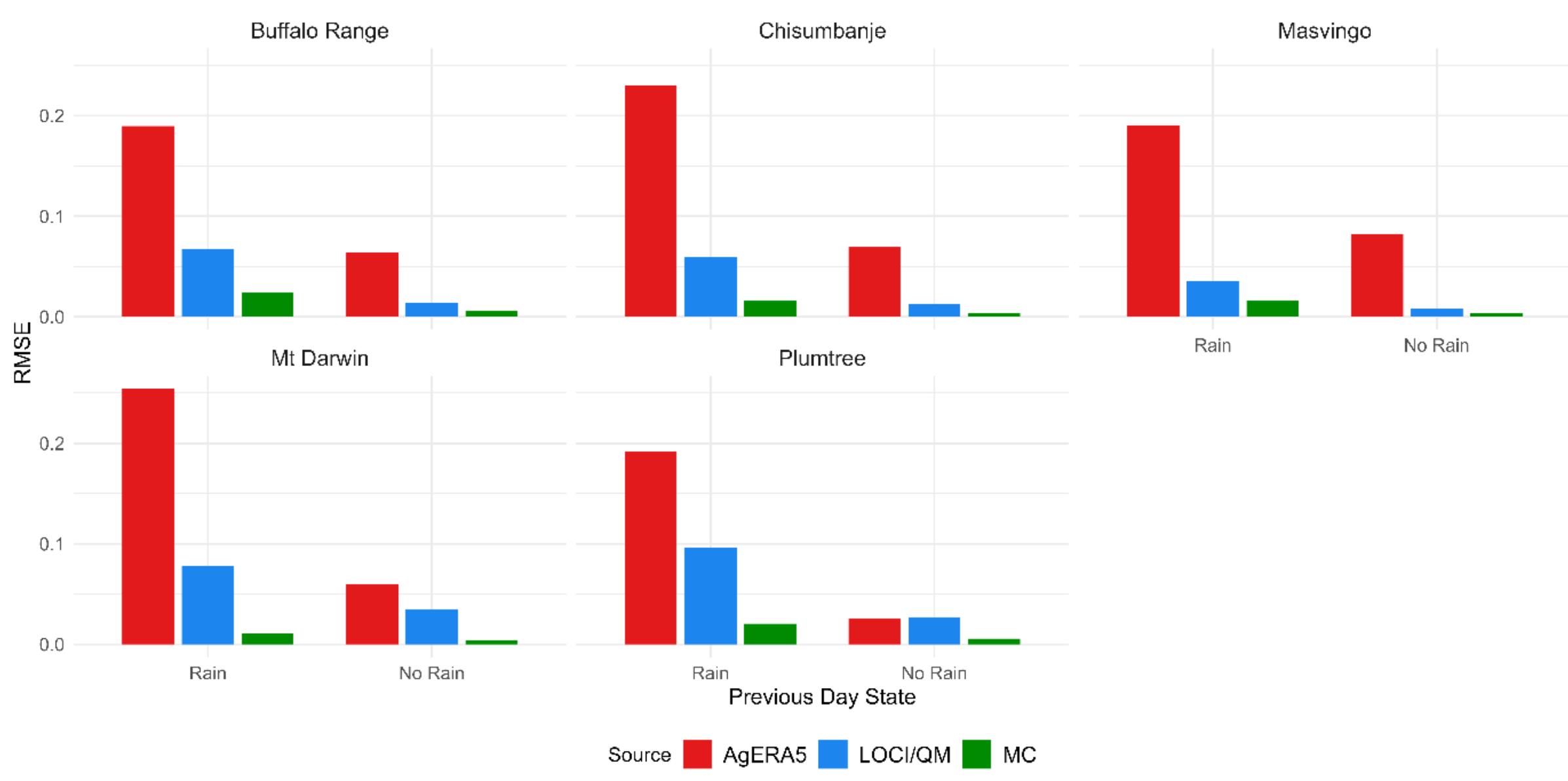

Figure 10: Root Mean Square Error (RMSE) values for Previous Day State=Rain ($RMSE_{curve}^{(W)}$) and Previous Day State=No Rain ($RMSE_{curve}^{(D)}$) for AgERA5, LOCI/QM, and MC compared with gauge values for the five stations.

### 5.2.5 Rainfall occurrence detection

Figure 11 shows the rain day probability of detection (POD), false alarm ratio (FAR), and Heidke skill score (HSS) at the five stations. AgERA5 had a high POD, typically >0.80 (Table S4). However, the FAR was also high (>0.50 at all stations). The HSS, combining skill balancing wet and dry day detection, misses, and false alarms, was typically <0.50.

Results for LOCI/QM and MC were comparable. The POD was lower than for AgERA5 but was generally above 0.50 (Table S4), and FAR below 0.50. HSS increased by approximately 0.06 on average compared with AgERA5.LOCI/QM had a slightly higher HSS compared with MC (<0.01 difference).

Plumtree was a consistent exception across all sources. POD was notably lower compared to the other stations (AgERA5 0.56; LOCI/QM and MC 0.38). FAR remained relatively high after bias correction, and HSS showed a small decrease for both LOCI/QM and MC compared to AgERA5.

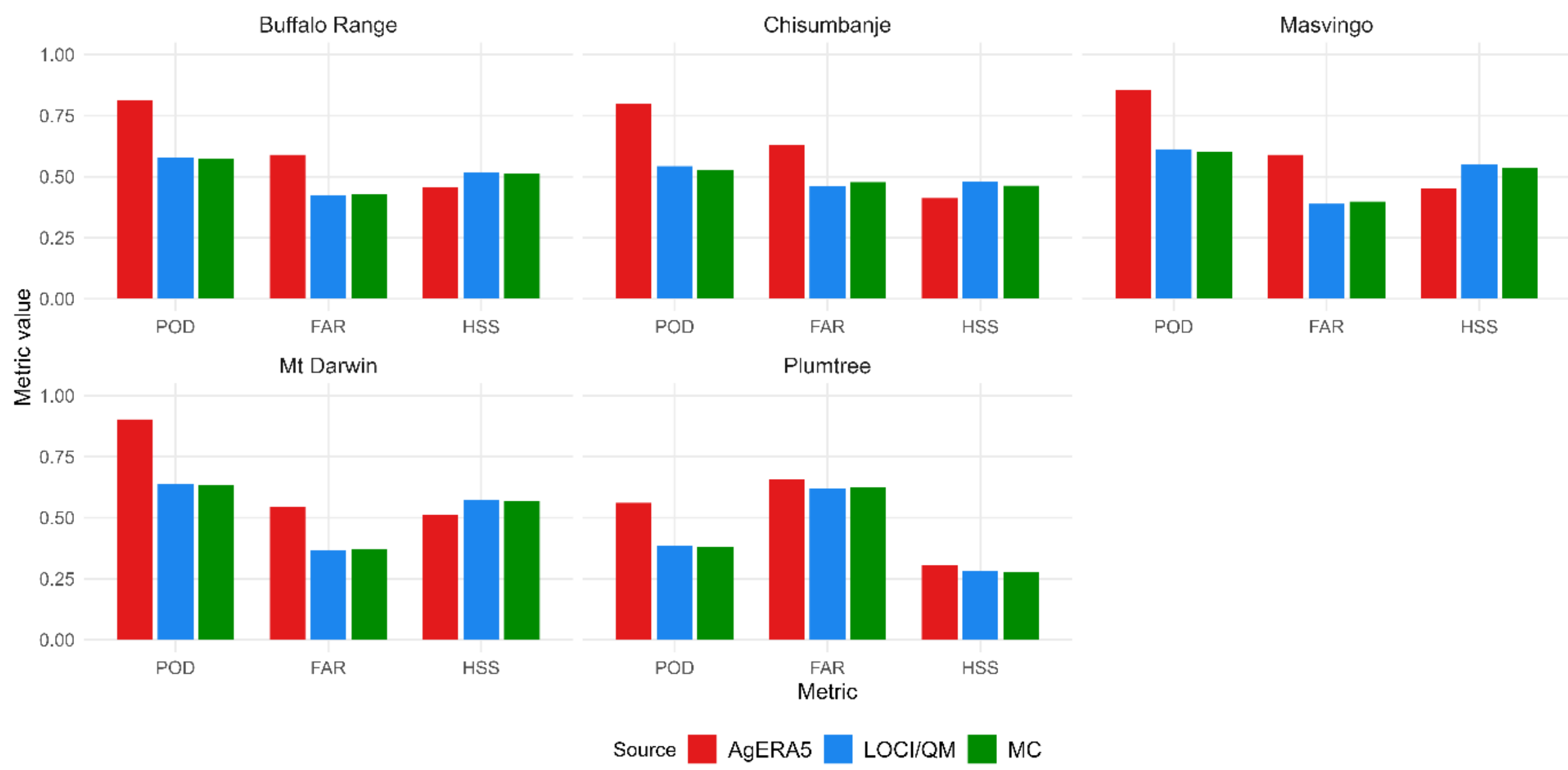

Figure 11: Probability of detection (POD), false alarm ratio (FAR), and Heidke skill score (HSS) for AgERA5, LOCI/QM and MC for the five stations.

## 5.3 Rainfall amounts

### 5.3.1 Climatology

**Mean monthly total rainfall**

AgERA5 captured the seasonal pattern of mean monthly rainfall, with a slight overestimation in most months, but underestimation of rainfall at the peak of the rainy season in January/February at four stations (Figure 12). The exception was Mt Darwin, where overestimation was greatest in December and January.

All four bias correction methods reduced the biases in AgERA5 and produced mean monthly totals closer to the gauge data in all months across the five stations. Differences between the methods were small, with mean monthly totals typically within 1-2 mm of each other.

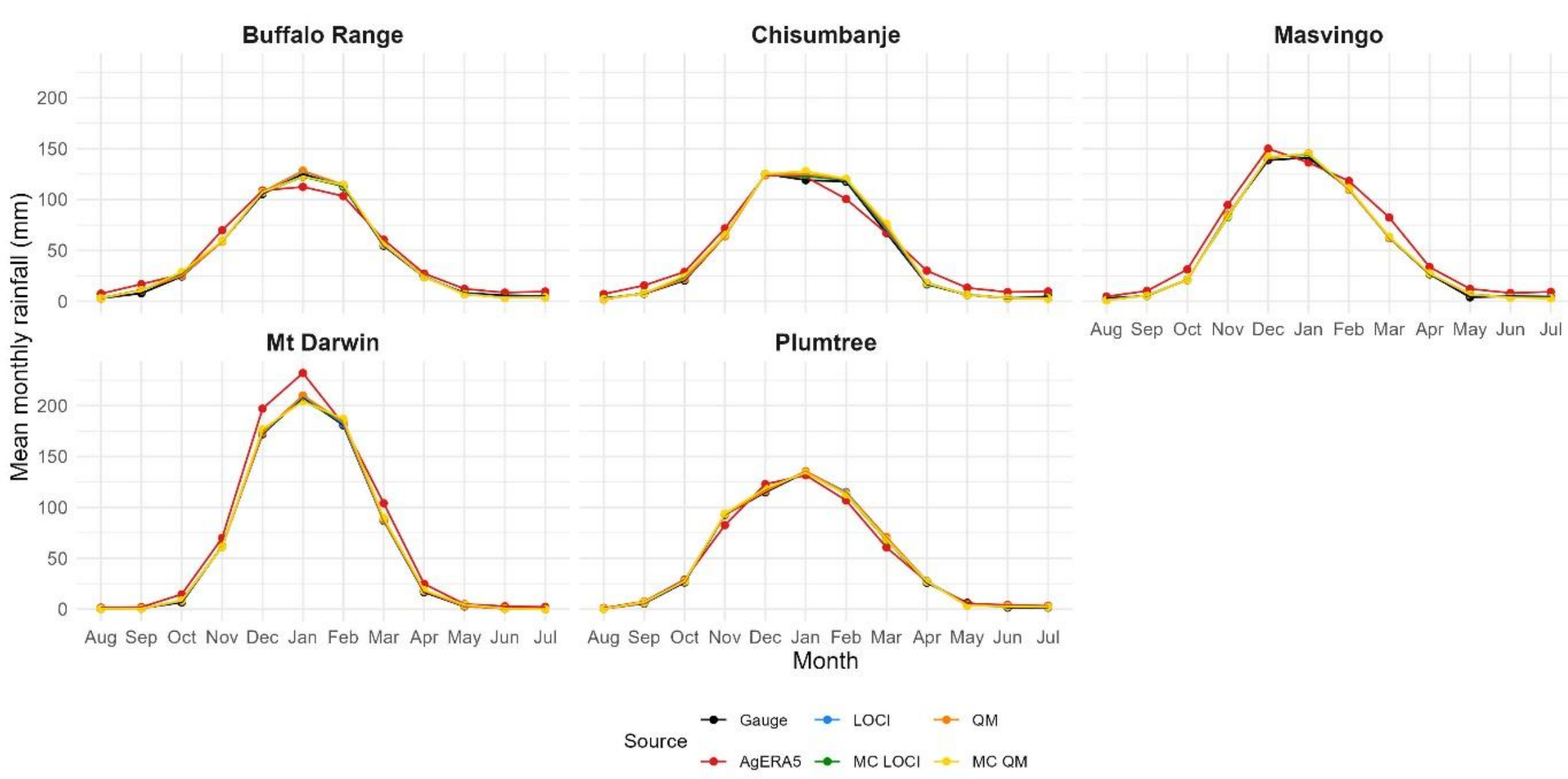

Figure 12: Mean monthly rainfall totals for Gauge, AgERA5, LOCI, MC LOCI, QM, and MC QM sources at the five stations.

**Mean rainfall per rain day**

AgERA5 captured the seasonal pattern of mean rainfall per rain day but consistently underestimated at all stations. All four bias correction methods reduced the bias of AgERA5, with smaller differences compared with the gauge data in almost all months across the five stations (Figure 13). At Buffalo Range, Masvingo, and Plumtree, all four methods performed similarly, with comparable RMSE values (Table S5). At Chisumbanje, LOCI and QM underestimated mean rainfall per rain day, during the transition between the rainy and dry seasons, more than their MC counterparts, but slightly outperformed MC LOCI and MC QM during the peak of the rainy season. At Mt Darwin, the four bias correction methods performed similarly in most months. Performance was worse from May to September where there were fewer rain days and hence mean rainfall per rain day was sensitive to individual large rainfall events. MC LOCI had slightly lower RMSE values than MC QM at all stations.

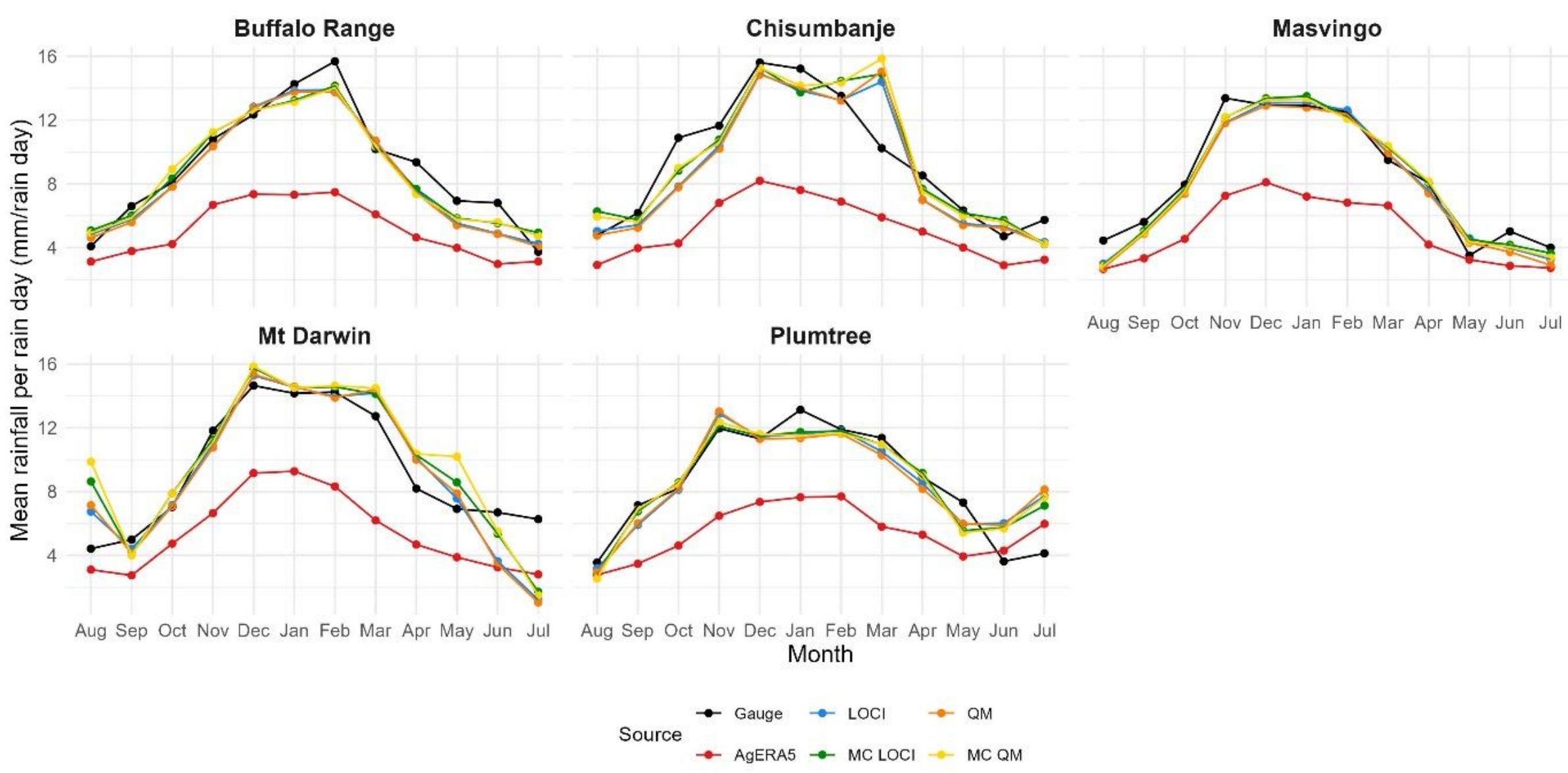

Figure 13: Mean monthly rainfall per rain day for Gauge, AgERA5, LOCI, MC LOCI, QM, and MC QM sources at the five stations.

**Maximum daily rainfall**

AgERA5 captured the seasonal pattern of maximum daily rainfall but consistently underestimated at each station, particularly during the peak rainy season (Figure 14). All four bias correction methods reduced the bias of AgERA5. During November to January, the two quantile mapping methods (QM and MC QM) produced larger maximum daily rainfall values, that were generally closer to the gauge data, than the LOCI and MC LOCI. MC QM produced slightly larger values than QM, usually closer to the gauge data, except at Mt Darwin where both overestimated maximum daily rainfall. LOCI and MC LOCI consistently underestimated maximum daily rainfall during the peak of the rainy season, but MC LOCI produced slightly larger values than LOCI that were closer to the gauge data (Figure 14).

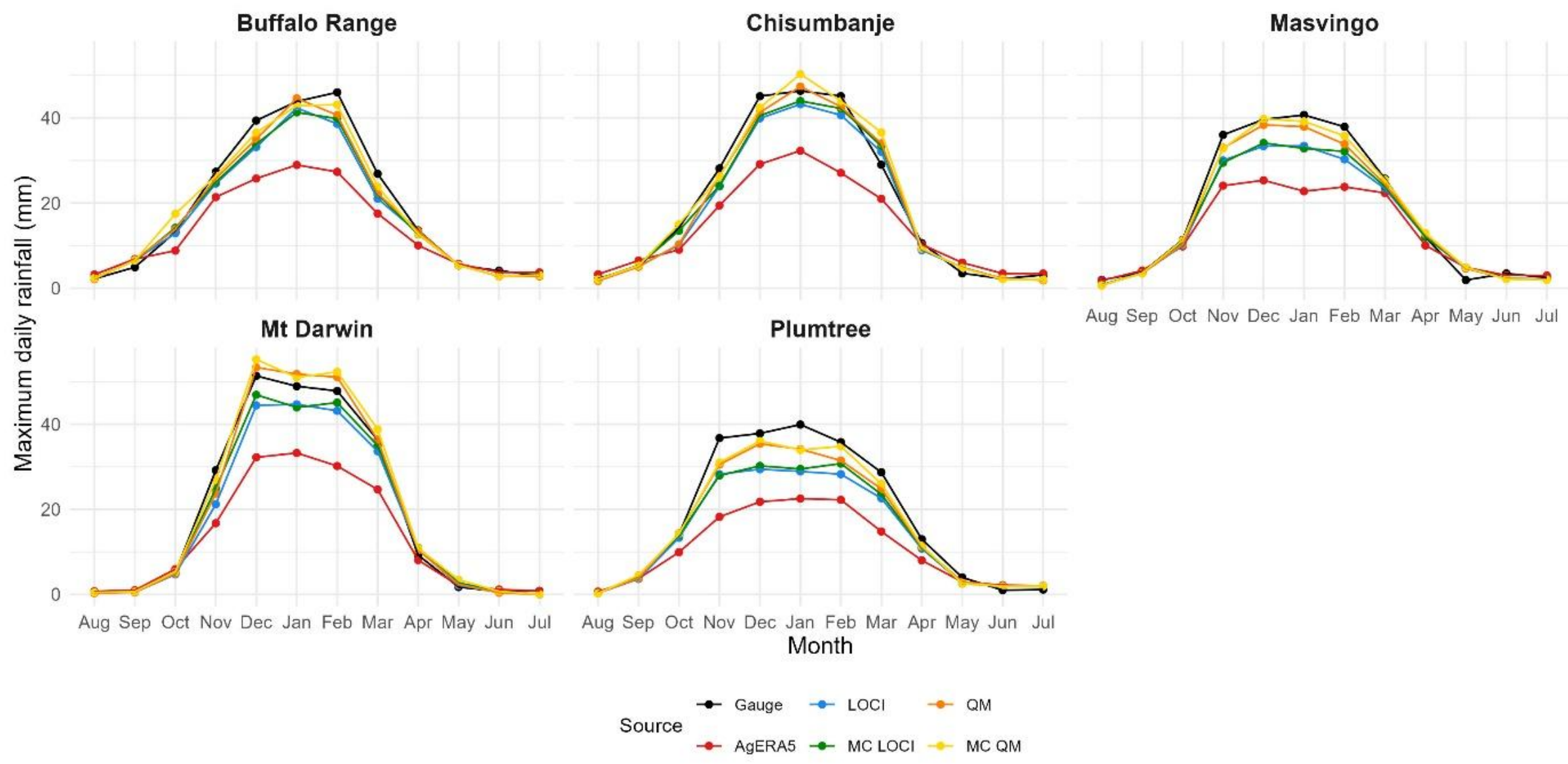

Figure 14: Monthly maximum daily rainfall for Gauge, AgERA5, LOCI, MC LOCI, QM, and MC QM sources at the five stations.

## 5.3.2 Annual summaries

**Annual total rainfall**

AgERA5 overestimated total annual rainfall on average, with ME values of 28.15-98.43 mm, except at Plumtree where there was a small underestimation (-10 mm) (Table S6). Correlation was moderate to strong positive at four stations and weak positive at Plumtree. rSD was close to 1 at Buffalo Range, Mt Darwin, and Plumtree, indicating a similar variability to the gauge values, with slightly lower variability at Chisumbaje and Masvingo.

All four bias correction methods reduced the average bias of AgERA5, with ME for all four bias correction methods <28.39 mm at all stations, slightly overestimating the gauge values. The lowest ME values were LOCI or MC LOCI at all stations. Correlations for the four bias correction methods were slightly lower than for AgERA5, but moderate to strong positive at the same four stations. Differences in correlation between the four methods were small, although the MC variants generally had slightly lower correlations than their non-MC equivalents.

The four bias correction methods produced more variable total annual rainfall values than AgERA5, with increased rSD values at all stations. This resulted in slightly more variable annual total values than the gauge data at Buffalo Range, Mt Darwin, and Plumtree, and more similar variability at the two other stations. The MC variants had slightly lower rSD values than their non-MC equivalents, usually resulting in rSD values closer to 1, and therefore more similar in variability to the gauge values. Time series graphs of the annual totals are shown in Figure S1.

**Annual mean rainfall per rain day**

AgERA5 consistently underestimated mean rainfall per rain day at all stations, with ME values of -6.06 to -4.75 mm (Table S6, Figure 15). Correlation was weak positive or negligible, and variability was substantially lower than the gauge data (rSD<1 at all stations).

All four bias correction methods reduced bias compared with AgERA5, with ME values of -0.26 to 1.50 mm across all stations and methods. Differences between the four methods were small, although the average bias was generally slightly larger for the MC variants. All four bias correction methods had increased rSD values compared to AgERA5, with values closer to 1 at all stations. The MC variants generally had slightly lower rSD values than their non-MC equivalents, usually resulting in rSD values closer to 1. Correlation remained weak positive to negligible, often lower than that of AgERA5. At Plumtree, where AgERA5 correlation was close to 0, this led to slightly negative correlation for all four bias correction methods.

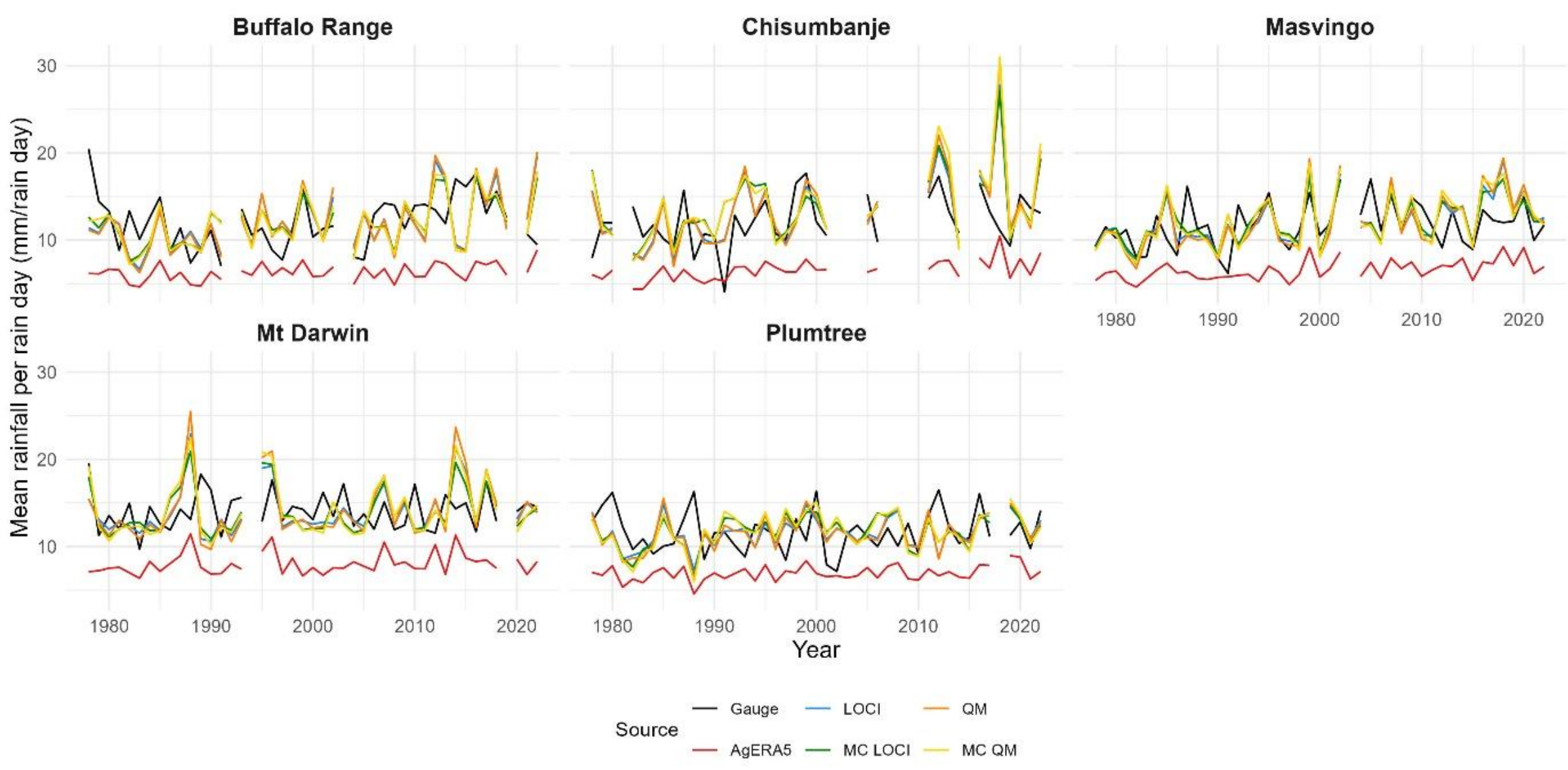

Figure 15: Annual mean rainfall per rain day for Gauge, AgERA5, LOCI, MC LOCI, QM, and MC QM at the five stations.

**Annual maximum rainfall**

AgERA5 consistently underestimated maximum daily rainfall, with ME values of -28.76 to -19.32 mm (Table S6). Correlation was weak positive or negligible across the stations. Variability was lower than the gauge values at three stations (rSD <1), with rSD values close to 1 at Chisumbanje and Mt Darwin.

The four bias correction methods generally reduced the average bias in maximum daily rainfall, though no method consistently outperformed the others across all stations. QM and MC QM mostly produced the largest extreme maximum values across the stations, sometimes exceeding any value in the gauge data by a large amount (Figure S2). ME values for MC LOCI and LOCI were similar, as were ME values for MC QM and QM. Correlation was slightly lower than AgERA5 for all four methods, with MC variants marginally lower than their non-MC counterparts. All four bias correction methods had increased rSD relative to AgERA5 and often closer to 1. However, when the rSD value for AgERA5 was already close to 1, the bias correction methods produced much more variability than the gauge data, particularly at Mt Darwin which had very high rSD values for QM and MC QM (2.38 and 2.24, respectively) (Table S6).

### 5.3.3 Seasonal distribution

To assess robustness across locations and validation periods, the mean and standard deviation of the RMSE metrics for the zero- and first-order Markov chain models of mean rainfall per rain day were calculated across the 20 station-validation block combinations and are presented in Table 11. This complements the station-level results presented in this section.

| Metric | AgERA5 | LOCI | MC LOCI | QM | MC QM |
|---|---|---|---|---|---|
| $RMSE_{curve}^{(0)}$ | 4.730 (1.003) | 1.950 (0.925) | 2.046 (0.923) | 2.172 (1.167) | 2.294 (1.280) |
| $RMSE_{curve}^{(D)}$ | 6.067 (1.127) | 2.656 (1.105) | 2.047 (0.916) | 3.019 (1.418) | 2.271 (1.243) |
| $RMSE_{curve}^{(W)}$ | 4.332 (1.385) | 3.160 (2.010) | 2.366 (1.072) | 3.904 (3.415) | 2.772 (1.667) |

Table 11: Mean (standard deviation) values of the Root Mean Square Error (RMSE) metrics for the zero- ($RMSE_{curve}^{(0)}$) and first-order ($RMSE_{curve}^{(D)}$, $RMSE_{curve}^{(W)}$) Markov chain models of mean rainfall per rain day across the 20 station-validation block combinations for AgERA5, LOCI, MC LOCI, QM and MC QM methods.

**Zero-order Markov chain models of mean rainfall per rain day**

The zero-order Markov chain model graphs in Figure 16 show the estimated mean rainfall per rain day for each day of the year from August to July, to align with the rainy season. AgERA5 captured the seasonal pattern but consistently underestimated mean rainfall at every station, with $RMSE_{curve}^{(0)}$ values of 3.96 to 5.28 mm (Table 12).

All four bias correction methods improved the estimates, reducing $RMSE_{curve}^{(0)}$ values and giving a closer estimation of the gauge data distribution across the five stations (Table 12). The station-validation block summary showed a similar pattern. All four bias correction methods produced comparable mean and standard deviation values for $RMSE_{curve}^{(0)}$, substantially improving on AgERA5 (Table 11).

The performance of the bias correction methods was similar to each other at most stations. The exception was Mt Darwin, where there were substantial differences in $RMSE_{curve}^{0}$ values, with LOCI and QM underestimating mean rainfall per rain day in the middle of the dry season where there was generally less data. At Buffalo Range, MC LOCI and MC QM had better estimate values at the peak of the rainy season than LOCI and QM.

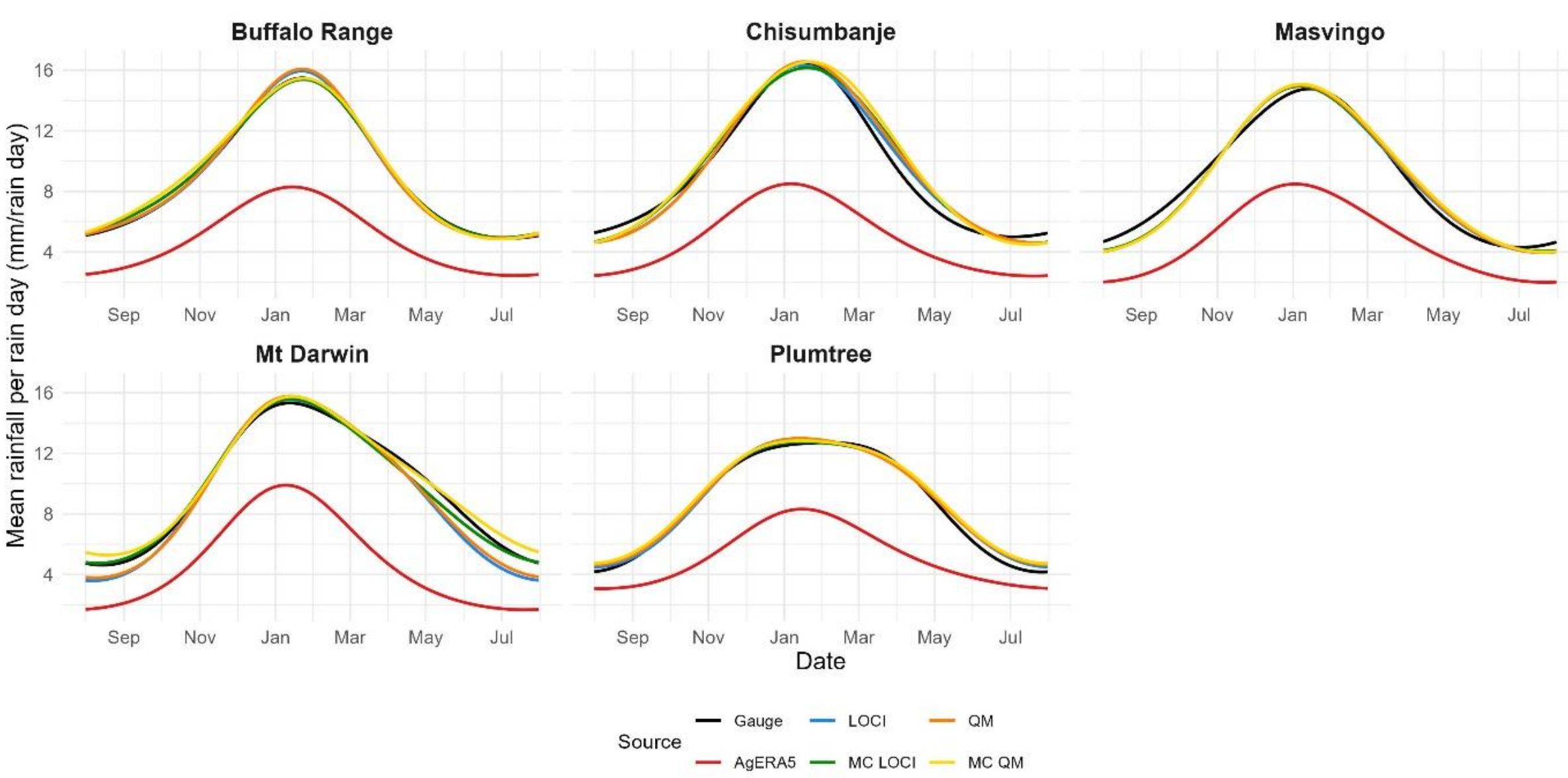

Figure 16: Zero-order Markov chain model curves showing the estimated mean rainfall per rain day for each day of the year, from August to July at the five stations, for Gauge, AgERA5, LOCI, and MC LOCI sources.

| Station | AgERA5 | LOCI | MC LOCI | QM | MC QM |
|---|---|---|---|---|---|
| Buffalo Range | 4.60 | 0.20 | **0.18** | 0.25 | 0.33 |
| Chisumbanje | 4.99 | **0.52** | 0.63 | 0.63 | 0.79 |
| Masvingo | 4.36 | 0.50 | **0.49** | 0.51 | 0.56 |
| Mt Darwin | 5.28 | 0.85 | **0.37** | 0.74 | 0.40 |
| Plumtree | 3.96 | **0.30** | 0.32 | 0.33 | 0.39 |

Table 12: Root Mean Square Error ($RMSE_{curve}^{(0)}$) values for mean rainfall per rain day for AgERA5, LOCI, MC LOCI, QM, and MC QM compared to the gauge data for the five stations.

**First-order Markov chain models of mean rainfall per rain day**

The first-order Markov chain model curves show the estimated mean rainfall per rain day, conditional on the state of the previous day (rain or no rain) for each day of the year, from August to July (Figure 17).

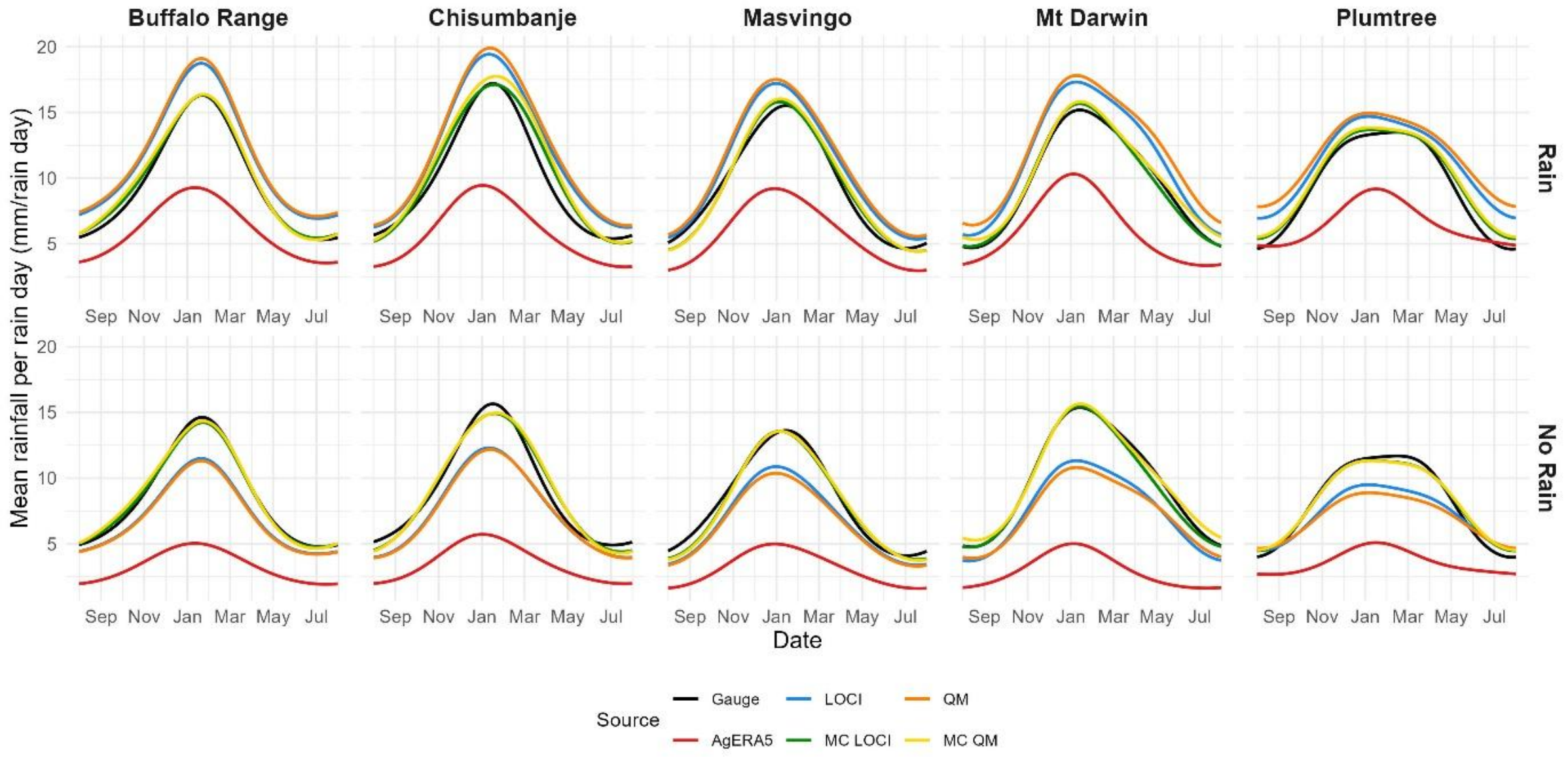

Figure 17: First-order Markov chain model curves showing the estimated mean rainfall per rain day with separate curves conditional on the state of the previous day (rain or no rain), for each day of the year, from August to July at the five stations, for Gauge, AgERA5, LOCI, QM, MC LOCI, and MC QM sources.

AgERA5 showed a similar seasonal pattern of mean rainfall per rain day to the gauge data for both previous day states across all stations, but with significant underestimation. In the peak of the rainy season, AgERA5 values were as low as half the gauge data when the previous day was wet, and as low as a third when the previous day was dry.

Figure 18 shows the $RMSE_{curve}^{(W)}$ and $RMSE_{curve}^{(D)}$ values for mean rainfall per rain day at each station. AgERA5 had the largest $RMSE_{curve}^{(W)}$ and $RMSE_{curve}^{(D)}$ values (3.7-4.6 mm) when the previous day was wet and values of 5.1-7.5 mm when the previous day was dry (Table S7).

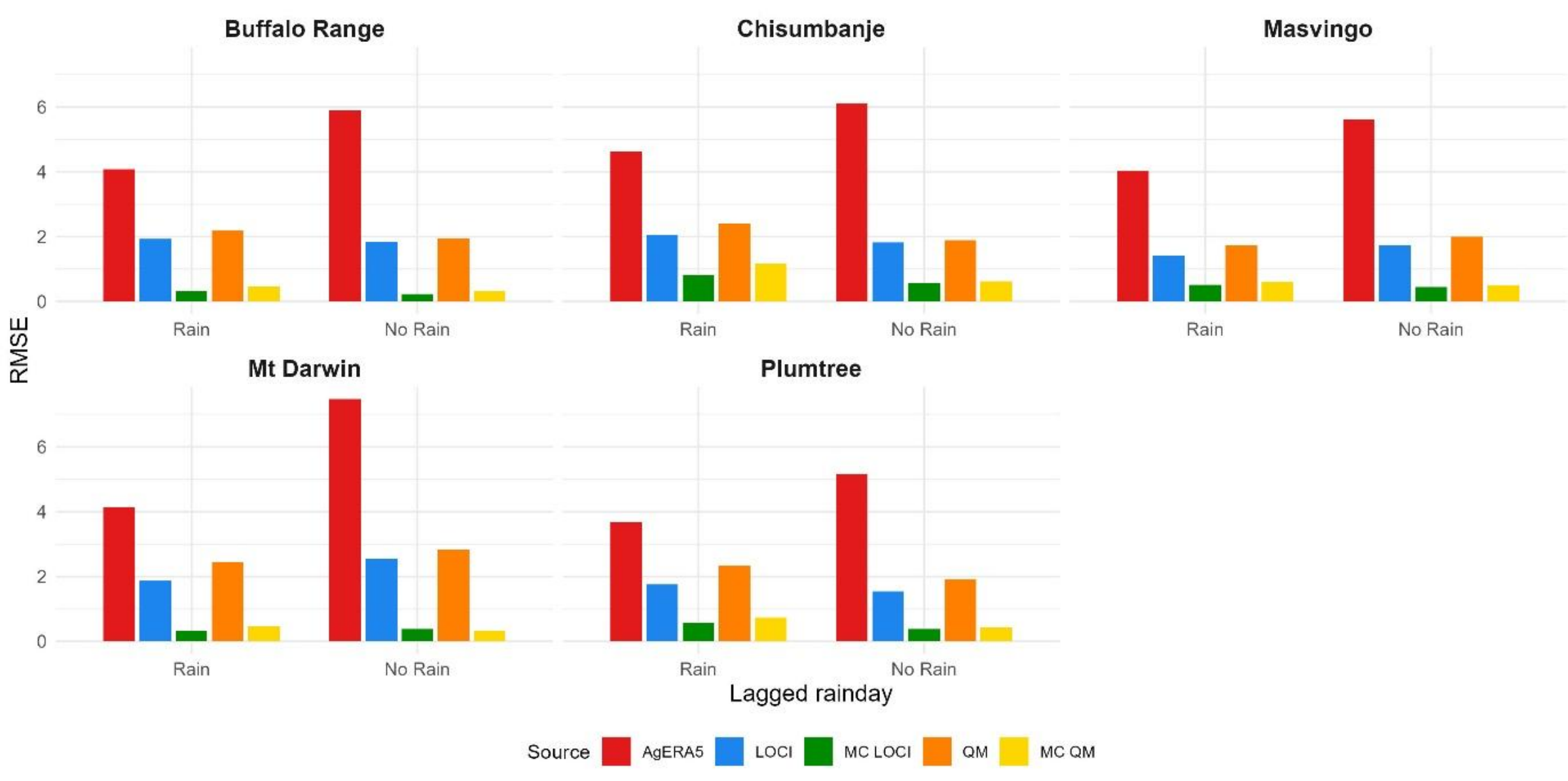

Figure 18: Root Mean Square Error ($RMSE_{curve}^{(W)}$ and $RMSE_{curve}^{(D)}$) values for mean rainfall per rain day for AgERA5, LOCI, MC LOCI, QM, and MC QM compared with the Gauge data for the five stations.

LOCI and QM reduced some of the biases in AgERA5 while maintaining a similar seasonal pattern to the gauge data (Figure 17). RMSE values were typically reduced by up to half for the wet state, and by up to two thirds for the dry state (Figure 18). However, both LOCI and QM overestimated mean rainfall per rain day when the previous day was wet and underestimated mean rainfall when the previous day was dry, with RMSE values approximately 2 mm in both cases (Table S7). The largest differences were typically during the peak of the rainy season when the previous day was dry (Figure 17).

MC LOCI and MC QM aligned closely with the gauge curves for both previous day states at all stations. Their $RMSE_{curve}^{(W)}$ and $RMSE_{curve}^{(D)}$ values were lower than AgERA5, LOCI, and QM's corresponding values for all stations, with values below 1 mm (Figure 18 and Table S7). Throughout the year, MC LOCI and MC QM curves had the closest estimation of mean rainfall per rain day. The difference between MC LOCI and MC QM was small, with MC QM having a slightly higher RMSE at most stations for both states. This pattern was also observed in the station-validation block summary, where both MC LOCI and MC QM achieved lower mean and standard deviation values for $RMSE_{curve}^{(W)}$and $RMSE_{curve}^{(D)}$than LOCI and QM (Table 11). MC LOCI had slightly lower mean and standard deviation values than MC QM for both previous day states.

### 5.3.4 Rainfall amount detection

Figure 19 shows the probability of detection (POD) and the overall Heidke Skill Score (HSS) of the daily rainfall values grouped into five categories of rainfall (no rain, light rain, moderate rain, heavy rain, and violent rain), as defined in Section 4.2.

All four bias correction methods increased the detection of dry days compared with AgERA5 across all stations. For light and moderate rain, AgERA5 had a higher POD than all bias correction methods, approximately double for light rain, with a slightly smaller increase in POD for moderate rain. For heavy and violent rain, all bias correction methods outperformed

AgERA5. AgERA5 showed a particularly low POD for violent rain, which all bias correction methods improved, although detection remained below 0.25 for all four bias correction methods at all stations, and as low as 0.08 at Plumtree. HSS, representing the skill over all five rainfall categories, was slightly higher for all four bias correction methods at all stations except Plumtree, where AgERA5 was the highest.

Across the four bias correction methods, dry day detection was similar at all stations. For other categories, QM and MC QM outperformed LOCI and MC LOCI for light rain at three stations, while LOCI and MC LOCI outperformed QM and MC QM for moderate rain at four stations. Heavy rain detection was similar for all four methods, whereas QM or MC QM performed best at each station for violent rain. Overall, HSS values were similar, although the non-MC methods generally outperformed their MC counterparts.

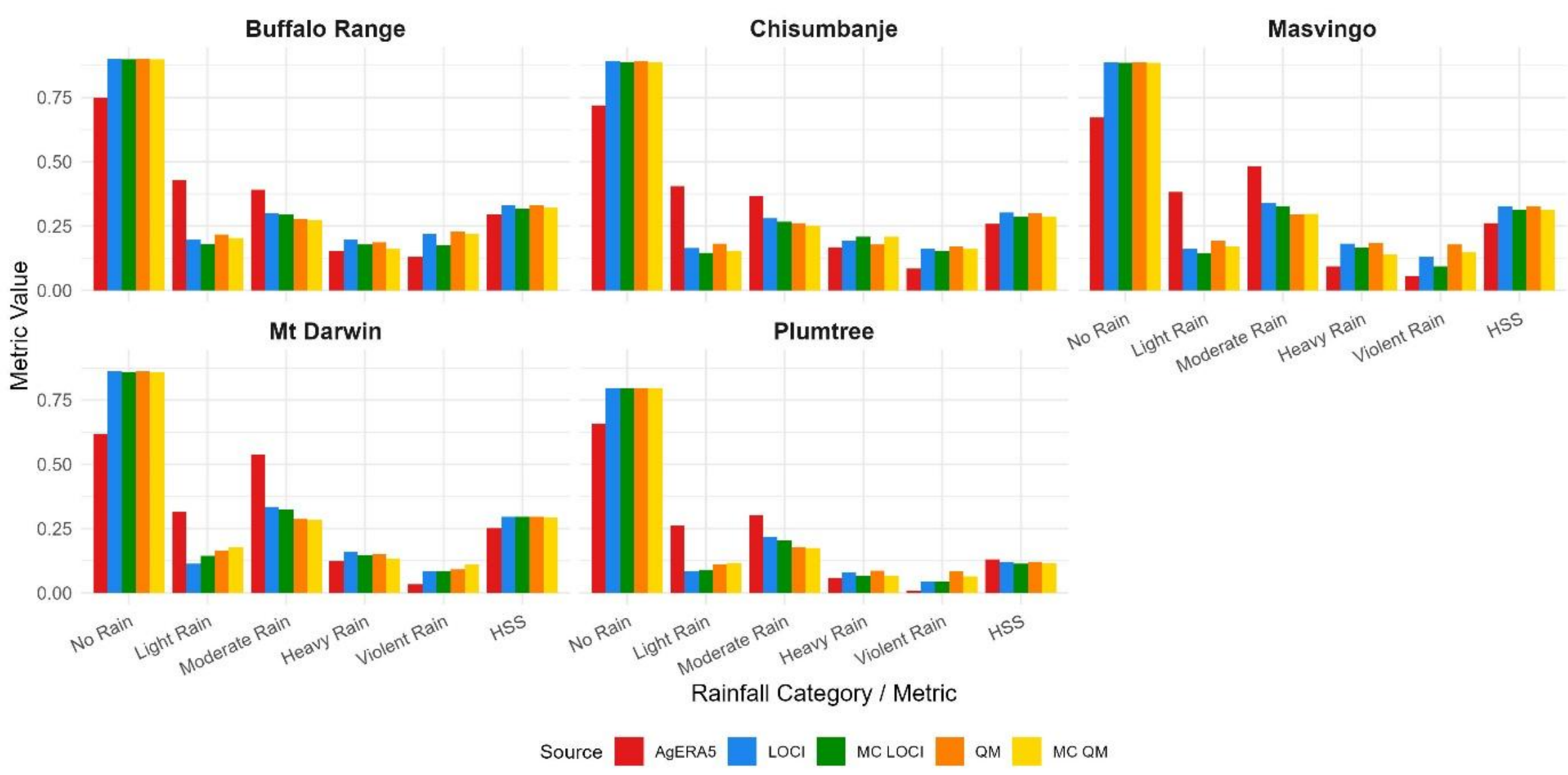

Figure 19: Probability of detection for five intervals of daily rainfall amounts, as defined in section 4.2, and overall Heidke Skill Score (HSS) of the daily rainfall values grouped into the five categories shown, for AgERA5, LOCI, MC LOCI, QM, and MC QM compared to the Gauge data at the five stations.

# 6 Discussion

## 6.1 Stability and robustness of the Markov chain bias correction method

The results of our study showed that the Markov chain bias correction calibration method accurately reproduced rainfall occurrence probabilities, as demonstrated by the close agreement between target and calibrated probabilities for all stations, months, and calibration blocks. The close matching of both conditional probabilities ($p_w$ and $p_d$) and the unconditional probability ($p_0$) confirmed the internal consistency of the method, and also showed that the method is stable across the five tested locations, suggesting that the cross-block calibration approach reduces any influence of non-stationarity in rain day thresholds over time. This consistent performance across four independent calibration blocks spanning 44 years highlights the robustness of the method under varying temporal conditions.

The largest deviations between target and calibrated probabilities were observed for $p_w$ during the dry season and transition months of October and April, attributable to the limited number of rain-after-rain days during these periods and hence the resulting small sample sizes for calibration. This finding could result in reduced accuracy in these periods, highlighting a potential limitation of the method when rainfall frequency is low. These findings suggest that refinement to the method may be beneficial to improve reliability where data availability is limited in general.

## 6.2 Performance of bias correction methods for rainfall occurrence

Our results showed that LOCI and QM (labelled LOCI/QM) and MC LOCI and MC QM (labelled MC) corrected the systematic overestimation of the number of rain days by AgERA5 on a seasonal and annual basis, with similar reductions in ME, while largely preserving the temporal variability, shown by similar annual correlation values. This is in agreement with previous studies in other geographic contexts, which have shown LOCI and QM to be effective at correcting systematic bias in rainfall frequency (Fang, Yang, Chen, & Zammit, 2015; Gudmundsson, Bremnes, Haugen, & Skaugen, 2012).

MC outperformed LOCI/QM for estimation of the annual longest dry spell and the distribution of wet and dry spells. LOCI/QM reduced the overestimation of dry spell frequency but tended to overcorrect, resulting in underestimation and still showed significant differences from observed distributions at several stations, as has been observed in other studies (Liu, Chen, Zhang, Xu, & Hui, 2020). MC provided the closest agreement with gauge data for both wet and dry spell distributions, with consistently lower K–S statistics and largely non-significant differences, indicating a more similar estimation of spell lengths to the gauge data. However, no method improved on the persistently weak correlations of AgERA5 for the annual longest dry spell metrics. This finding indicates that capturing the timing of long dry periods was not substantially improved by either bias correction method, likely limited by the inherent ability of AgERA5 to accurately estimate the timing of rainfall events.

On a seasonal basis, LOCI/QM and MC reduced AgERA5's bias in rainfall occurrence, while maintaining a seasonal distribution similar to the gauge data, as shown in the zero-order Markov chain curves. However, for first-order Markov chain curves, which divided rainfall occurrence according to whether or not the previous day was a rain day, LOCI/QM overestimated the probability of rainfall when the previous day was a rain day, and hence gave higher probabilities of wet spells continuing (rainfall persistence), and underestimated rain when the previous day had no rain day, giving lower probabilities of dry spells ending (rainfall onset). MC outperformed LOCI/QM on both first-order Markov chain curves, substantially reducing RMSE values at all stations and more accurately capturing the probabilities related to both rainfall persistence and rainfall onset.

The station-validation block summaries show that the MC methods achieved lower mean values for the spell length and first order Markov chain metrics and also showed lower variability. This indicates that the improvements were generally robust and consistent across locations and validation periods.

These findings demonstrate that the MC based approach improved the representation of temporal rainfall structure beyond the single threshold based approach of LOCI and QM at the five locations evaluated. The MC based approach achieved this improvement while maintaining a similar performance to LOCI/QM on standard rainfall occurrence characteristics.

The rainfall occurrence detection results showed that AgERA5 had a high probability of detection, but also a high false alarm ratio, resulting in moderate overall skill, with HSS values generally <0.50. LOCI/QM and MC improved the balance between correctly detecting rain days and false alarms, leading to higher overall skill at most stations. The similar performance of LOCI/QM and MC suggests that while the MC based approach enhanced overall temporal characteristics of rainfall occurrence, this approach does not seem to provide any additional benefit for daily occurrence detection skill.

The similar detection skill between the methods is not surprising given their similarity in the use of threshold based approaches to define rain days. This indicates that the methods are potentially limited by the underlying ability of AgERA5 to detect rainfall events observed by the gauge data, as has been described by others (Xiao, Zou, Xia, Yang, & Yao, 2022).

## 6.3 Performance of bias correction methods for rainfall amounts

All four bias correction methods substantially reduced the bias in total rainfall at both monthly climatology and annual scales. MC LOCI had the lowest ME value at three stations, and LOCI at the other two stations, suggesting that the mean based methods provided a better representation of monthly and annual total rainfall than the quantile mapping based methods. However, analysis across more locations, climate conditions and gridded products would be needed to assess whether the improvements of MC LOCI over LOCI for total rainfall are consistent.

The consistent underestimation of mean rainfall per rain day by AgERA5 was primarily a consequence of its overestimation of rain day frequency. All four bias correction methods effectively reduced this bias and improved inter-annual variability of mean rainfall per rain day. MC LOCI had the lowest RMSE value for the monthly climatology at four stations, with QM or MC QM having the highest across the stations.

The quantile mapping based methods generally gave better representation of extreme rainfall, with maximum daily rainfall closer to observations than the mean based methods. This behaviour is widely reported in previous studies (Teutschbein & Seibert, 2012; Gudmundsson, Bremnes, Haugen, & Skaugen, 2012; Themeßl, Gobiet, & Leuprecht, 2011), which show that distribution based methods more effectively represent extreme rainfall. However, all methods reduced the underestimation of maximum rainfall by AgERA5. The quantile mapping based methods, however, seemed to have a greater potential for occasional generation of unrealistically large rainfall values beyond the gauge data values, and much larger variability in maximum rainfall than the gauge data. This behaviour is a recognised limitation of quantile mapping methods (Maraun, 2016).

Correlations of annual summaries generally decreased for all four methods relative to AgERA5, indicating that while average biases in rainfall magnitude were corrected, the timing, particularly of large rainfall events, might not be substantially improved by these bias correction methods. This highlights a general limitation of bias correction methods and suggests that additional information may be needed to improve the timing of rainfall events.

MC variants tended to produce slightly less variable annual summaries than their non-MC counterparts, and rSD values closer to 1 at most stations. This finding could indicate that the conditional adjustment of amounts based on the state of the previous day, might produce values that result in summaries with more realistic variability.

All four bias correction methods reduced the significant underestimation in mean rain per rain day by AgERA5 on a seasonal basis, as shown by the zero-order Markov chain curves. In the first-order Markov chain curves, LOCI and QM reduced some of the bias of AgERA5, but consistently overestimated mean rainfall per rain day when the previous day was wet, and underestimated mean rainfall when the previous day was dry. Both MC LOCI and MC QM closely agreed with the gauge curves and consistently had the lowest $RMSE_{curve}^{(W)}$ and $RMSE_{curve}^{(D)}$ values, and little difference between the two methods. The station-validation block summaries were consistent with this finding, with both MC LOCI and MC QM achieving lower mean and variability in $RMSE_{curve}^{(W)}$ and $RMSE_{curve}^{(D)}$ values than LOCI and QM, indicating that the improvements were generally robust across locations and validation periods. These findings indicate that incorporating temporal dependence through an MC based approach may improve the representation of rainfall intensity, conditional on the previous day's state. The MC based approach could provide a more accurate estimate of both the persistence of rainfall amounts following wet days and the average rainfall amounts associated with transitions from dry to wet conditions.

AgERA5's lower POD of dry days was mainly because of its overestimation of rainy days, which was largely corrected by all four bias correction methods. All four bias correction methods, however, had a lower POD for light rain and moderate rain than AgERA5, indicating that many of AgERA5's correctly categorised lower intensity rainfall amounts were recategorised by the bias correction. The largest reduction in POD was for light rain, where some of AgERA5's correct values were lost because of the threshold based approach of bias correction, reflecting a trade-off between correcting wet day frequency and maintaining detection of lower intensity rainfall. The four bias correction methods improved the detection of heavy and violent rainfall, although skill remained low, indicating persistent challenges in estimating high intensity rainfall. QM and MC QM had the highest detection of violent rain, indicating their ability to better adjust for extreme events than mean based methods, as observed in previous studies (Teutschbein & Seibert, 2012; Themeßl, Gobiet, & Leuprecht, 2011). The small increase in overall combined skill shows that these bias correction methods are limited in their ability to correct for the precise timing and intensity of rainfall events not already captured by the original data source. This finding indicates that incorporating additional data sources may be required to substantially improve overall categorical detection skill.

## 6.4 Future work and possible extensions

While this study presents several promising results, some limitations have been identified which could be addressed in future work. The evaluation was based on five station locations across two climatic zones in Zimbabwe, using one gridded dataset for bias correction. To assess the broader applicability of the bias correction methods, further evaluation is needed across a greater number of locations, climate zones and gridded datasets.

Future work should therefore evaluate the proposed methods under a wider range of climatic conditions to confirm that the stability and robustness observed in our study are maintained under different rainfall patterns and datasets. Some variation in performance metrics was seen across the five sites and two climatic zones of this study. For example, at Mt Darwin, in a humid subtropical zone, the distribution of dry spells was not well matched by the MC methods, compared to the other four sites in the semi-arid climate zone. At Plumtree, rainfall occurrence and amount detection were not substantially improved by any bias correction method. Variation across locations should be investigated further, and evaluation in regions with dense station networks would also allow for incorporating spatial dependence into parameter calibration, potentially improving consistency across locations.

In geographic contexts where rainfall persistence shows longer memory than the previous day, it could be beneficial to extend the approach to higher order Markov chain models. Higher order models would increase the number of parameters, however, potentially creating calibration challenges, particularly in lower rainfall periods. Hybrid-order Markov chain models offer a potential compromise by reducing the number of previous states while retaining additional memory, and have been proposed for use in specific geographic contexts (Stern & Coe, 1984; Wilks, 1999a). While monthly calibration with a single dry season was adopted in this study, alternative approaches have been used, such as a 61-day moving window period (Themeßl, Gobiet, & Leuprecht, 2011), which could improve parameter stability in rainfall limited periods. A further possible compromise could be to apply lower-order Markov chain models during dry seasons when rainfall variability is limited, and higher-order structures during the rainy season. These adaptive approaches would require careful testing to ensure stability and consistency.

Some aspects of rainfall estimation were not substantially improved by any of the bias correction methods, including rainfall amount detection and annual correlations, particularly for extreme rainfall events. This finding suggests that the bias correction methods are limited by the inherent skill of the underlying model data. Incorporating auxiliary data, such as moisture transport, sea surface temperatures or wind fields, could prove beneficial to better represent the timing and intensity of rainfall events, as these variables have been shown to influence rainfall variability and the development of extreme rainfall events in Southern Africa (Rapolaki, Blamey, Hermes, & Reason, 2021; Williams, Kniveton, & Layberry, 2007; Jury, 1996). Hybrid quantile mapping methods, which apply different correction mappings for the lower and upper quantiles, have also been shown to improve the representation of extremes (Holthuijzen, Beckage, Clemins, Higdon, & Winter, 2022) and could be incorporated into the proposed bias correction methods.

Finally, the proposed bias correction methods are fully deterministic, providing a single corrected rainfall realisation at each location. Extending the approach to stochastic bias correction, as suggested by Maraun (2016), or integrating it within weather generator frameworks, could enable probabilistic simulations that would allow for quantifying uncertainty and estimation of rainfall related risks.

# 7 Conclusions

In this study, we developed and evaluated a Markov chain based approach to bias correction by integrating a first-order Markov chain into the local intensity scaling (LOCI) and quantile mapping (QM) methods, resulting in two new methods: Markov chain local intensity scaling (MC LOCI) and Markov chain quantile mapping (MC QM). The aim of this approach was to improve the representation of the temporal structure of rainfall while preserving other rainfall characteristics captured by the original methods.

The original and proposed methods were applied to AgERA5 reanalysis rainfall data at five locations in Zimbabwe, bias corrected against long term gauge data records. The evaluation results showed that the Markov chain approach was stable and robust across the five stations locations, months, and calibration blocks. The proposed methods (MC LOCI and MC QM) provided a more accurate representation of rainfall persistence, onset, wet and dry spell distributions, and rainfall amounts during wet spells and in transitions from dry to wet conditions, compared to their respective original methods (LOCI and QM). Furthermore, the Markov chain methods maintained similar improvements in overall rain day frequency, and mean and total rainfall as the original methods. However, the proposed methods showed little improvement in event timing, and the detection of daily rainfall amounts, particularly for extreme events.

These results demonstrate the potential of incorporating temporal structure via a Markov chain integration into bias correction methods for applications such as crop simulation modelling, where the sequencing and persistence of rainfall strongly influence crop growth, soil moisture, and yield outcomes. The proposed methods could also be beneficial for hydrological modelling, drought monitoring, and water resources management, which rely on accurate representation of rainfall persistence, and wet and dry spells.

Priorities for future work include evaluation across a wider range of regions, climate zones and gridded datasets to assess results in broader contexts, as well as extensions to the approach, such as higher-order or hybrid Markov chains. Incorporating additional climatic variables, or hybrid methods, may also be beneficial to improve the timing of rainfall events and better detection of extreme rainfall.

# 8 References

Acharya, S. C., Nathan, R., Wang, Q. J., Su, C. H., & Eizenberg, N. (2019). An evaluation of daily precipitation from a regional atmospheric reanalysis over Australia. *Hydrology and Earth System Sciences, 23*(8), 3387-3403. doi:https://doi.org/10.5194/hess-23-3387-2019

Ailliot, P., Allard, D., Monbet, V., & Naveau, P. (2015). Stochastic weather generators: an overview of weather type models. *Journal de la société française de statistique, 156*(1), 101-115.

Amare, M., McKay, A., Tarp, F., & Barrett, C. B. (2018). Rainfall shocks and agricultural productivity: Implication for rural household consumption. *Agricultural Systems, 166*, 79–89. doi:https://doi.org/10.1016/j.agsy.2018.07.014

Boogaard, H., Schubert, J., De Wit, A., Lazebnik, J., Hutjes, R., & Van der Grijn, G. (2020). Agrometeorological indicators from 1979 to present derived from reanalysis. *Copernicus Climate Change Service (C3S) Climate Data Store (CDS)*. doi:https://doi.org/10.24381/cds.6c68c9bb

Burden, R. L., & Faires, J. D. (2011). *Numerical Analysis* (9th ed.). Boston: Brooks Cole.

Cannon, A. J., Sobie, S. R., & Murdock, T. Q. (2015). Bias Correction of GCM Precipitation by Quantile Mapping: How Well Do Methods Preserve Changes in Quantiles and Extremes? *Journal of Climate, 28*, 6938–6959. Retrieved from https://doi.org/10.1175/JCLI-D-14-00754.1.

Chen, D., & Chen, H. W. (2013). Using the Köppen classification to quantify climate variation and change: An example for 1901–2010. *Environmental Development, 6*, 69-79. doi:https://doi.org/10.1016/j.envdev.2013.03.007

da Silva Jale, J., Júnior, S. F., Xavier, É. F., Stošić, T., Stošić, B., & Ferreira, T. A. (2019). Application of Markov chain on daily rainfall data in Paraíba-Brazil from 1995-2015. *Acta Scientiarum. Technology, 41*, e37186. doi:https://doi.org/10.4025/actascitechnol.v41i1.37186

Ebita, A., Kobayashi, S., Ota, Y., Moriya, M., Kumabe, R., Onogi, K., & Ishimizu, T. (2011). The Japanese 55-year reanalysis “JRA-55”: an interim report. *Sola, 7*, 149-152. doi:https://doi.org/10.2151/sola.2011-038

Eyring, V., Bony, S., Meehl, G. A., Senior, C. A., Stevens, B., Stouffer, R. J., & Taylor, K. E. (2016). Overview of the Coupled Model Intercomparison Project Phase 6 (CMIP6) experimental design and organization. *Geoscientific Model Development, 9*(5), 1937–1958. doi:https://doi.org/10.5194/gmd-9-1937-2016

Fang, G. H., Yang, J., Chen, Y. N., & Zammit, C. (2015). Comparing bias correction methods in downscaling meteorological variables for a hydrologic impact study in an arid area in China. *Hydrology and Earth System Sciences, 19*(6), 2547-2559. doi:https://doi.org/10.5194/hess-19-2547-2015

Frischen, J., Meza, I., Rupp, D., Wietler, K., & Hagenlocher, M. (2020). Drought risk to agricultural systems in Zimbabwe: A spatial analysis of hazard, exposure, and vulnerability. *Sustainability, 12*(3), 752. doi:https://doi.org/10.3390/su12030752

Froidurot, S., & Diedhiou, A. (2017). Characteristics of wet and dry spells in the West African monsoon system. *Atmospheric Science Letters, 18*(3), 125-131. doi:https://doi.org/10.1002/asl.734

Funk, C., Peterson, P., Landsfeld, M., Pedreros, D., Verdin, J., Shukla, S., . . . Michaelsen, J. (2015). The climate hazards infrared precipitation with stations - a new environmental record for monitoring extremes. *Scientific Data, 2*(1), 1-21. doi:https://doi.org/10.1038/sdata.2015.66

Gabriel, K., & Neumann, J. (1962). A Markov chain model for daily rainfall occurrence at Tel Aviv. *Quarterly Journal of the Royal Meteorological Society, 88*(375), 90-95. doi:https://doi.org/10.1002/qj.49708837511

Giorgi, F., & Gutowski Jr, W. J. (2015). Regional dynamical downscaling and the CORDEX initiative. *Annual Review of Environment and Resources, 40*, 467-490. doi:https://doi.org/10.1146/annurev-environ-102014-021217

Gudmundsson, L., Bremnes, J. B., Haugen, J. E., & Skaugen, T. E. (2012). Downscaling RCM precipitation to the station scale using quantile mapping–a comparison of methods. *Hydrology & Earth System Sciences Discussions, 9*(5), 6185–6201. doi:https://doi.org/10.5194/hessd-9-6185-2012

He, X., Sonnenborg, T. O., Refsgaard, J. C., Vejen, F., & Jensen, K. H. (2013). Evaluation of the value of radar QPE data and rain gauge data for hydrological modeling. *Water Resources Research, 49*, 5989–6005. doi:https://doi.org/10.1002/wrcr.20471

Hersbach, H., Bell, B., Berrisford, P., Hirahara, S., Horányi, A., Muñoz-Sabater, J., . . . Thépaut, J.-N. (2020). The ERA5 global reanalysis. *Quarterly Journal of the Royal Meteorological Society, 146*(730), 1999-2049. doi:https://doi.org/10.1002/qj.3803

Holthuijzen, M., Beckage, B., Clemins, P. J., Higdon, D., & Winter, J. M. (2022). Robust bias-correction of precipitation extremes using a novel hybrid empirical quantile-mapping method: Advantages of a linear correction for extremes. *Theoretical and Applied Climatology, 149*(1), 863-882. doi:https://doi.org/10.1007/s00704-022-04035-2

Huang, N., & Ma, C. (2014). Convergence Analysis and Numerical Study of a Fixed-Point Iterative Method for Solving Systems of Nonlinear Equations. *The Scientific World Journal*, 789459. doi:https://doi.org/10.1155/2014/789459

Huffman, G. J., Bolvin, D. T., Braithwaite, D., Hsu, K., Joyce, R., Xie, P., & Yoo, S.-H. (2019). *NASA global precipitation measurement (GPM) integrated multi-satellite retrievals for GPM (IMERG).* Retrieved from chrome-extension://efaidnbmnnnibpcajpcglclefindmkaj/https://gpm.nasa.gov/sites/default/files/document_files/IMERG_ATBD_V06.pdf

Hyndman, R. J., & Fan, Y. (1996). Sample Quantiles in Statistical Packages. *The American Statistician, 50*(4), 361-365. doi:https://doi.org/10.1080/00031305.1996.10473566

Ines, A. V., & Hansen, J. W. (2006). Bias correction of daily GCM rainfall for crop simulation studies. *Agricultural and forest meteorology, 138*(1-4), 44-53. doi:https://doi.org/10.1016/j.agrformet.2006.03.009

IPCC. (2022). *Climate Change 2022: Impacts, Adaptation and Vulnerability.* Cambridge University Press. doi:10.1017/9781009325844.

Jimoh, O., & Webster, P. (1996). The optimum order of a Markov chain model for daily rainfall in Nigeria. *Journal of Hydrology, 185*(1), 45-69. doi:https://doi.org/10.1016/S0022-1694(96)03015-6

Jury, M. R. (1996). Regional teleconnection patterns associated with summer rainfall over South Africa, Namibia and Zimbabwe. *International Journal of Climatology: A Journal of the Royal Meteorological Society, 16*(2), 135-153. doi:https://doi.org/10.1002/(SICI)1097-0088(199602)16:2%3C135::AID-JOC4%3E3.0.CO;2-7

Kaptué, A. T., Hanan, N. P., Prihodko, L., & Ramirez, J. (2015). Spatial and temporal characteristics of rainfall in Africa: Summary statistics for temporal downscaling. *Water Resources Research, 51*, 2668–2679. doi:https://doi.org/10.1002/2014WR015918

Liu, H., Chen, J., Zhang, X.-C., Xu, C.-Y., & Hui, Y. (2020). A Markov chain-based bias correction method for simulating the temporal sequence of daily precipitation. *Atmosphere, 11*(1), 109. doi:https://doi.org/10.3390/atmos11010109

Maidment, R. I., Grimes, D., Black, E., Tarnavsky, E., Young, M., Greatrex, H., . . . Alcántara, E. M. (2017). A new, long-term daily satellite-based rainfall dataset for operational monitoring in Africa. *Scientific Data, 4*, 170063. doi:https://doi.org/10.1038/sdata.2017.63

Mapurisa, B., & Chikodzi, D. (2014). An Assessment of Trends of Monthly Contributions to Seasonal Rainfall in South-Eastern Zimbabwe. *American Journal of Climate Change, 3*, 50-59. doi:https://doi.org/10.4236/ajcc.2014.31005

Maraun, D. (2016). Bias correcting climate change simulations-a critical review. *Current Climate Change Reports, 2*(4), 211-220. doi:https://doi.org/10.1007/s40641-016-0050-x

Mazvimavi, D. (2010). Investigating changes over time of annual rainfall in Zimbabwe. *Hydrology and Earth System Sciences, 14*, 2671–2679. doi:https://doi.org/10.5194/hess-14-2671-2010

McBride, C. M., Kruger, A. C., & Dyson, L. (2022). Changes in extreme daily rainfall characteristics in South Africa: 1921–2020. *Weather and Climate Extremes*.

Meteorological Services Department of Zimbabwe. (2025, 6 17). *Drought Occurrence in Zimbabwe*. Retrieved 3 25, 2026, from Meteorological Services Department oF Zimbabwe: https://www.weatherzw.org.zw/news/drought-occurrence-in-zimbabwe/

Muñoz-Sabater, J., Dutra, E., Agustí-Panareda, A., Albergel, C., Arduini, G., Balsamo, G., & Thépaut, J. N. (2021). ERA5-Land: A state-of-the-art global reanalysis dataset for land applications. *Earth System Science Data Discussions, 13*(9), 4349–4383. doi:https://doi.org/10.5194/essd-13-4349-2021

Mushawemhuka, W. J., Fitchett, J. M., & Hoogendoorn, G. (2021). Towards quantifying climate suitability for Zimbabwean nature-based tourism. *South African Geographical Journal, 103*(4), 443-463. doi:https://doi.org/10.1080/03736245.2020.1835703

Panofsky, H. A., & Brier, G. W. (1968). *Some applications of statistics to meteorology.* The Pennsylvania State University Press.

Parsons, D., Stern, D., Ndanguza, D., Sylla, M. B., Musyoka, J., Bagiliko, J., . . . Dorward, P. (2026). Evaluating Satellite and Reanalysis Rainfall Estimates for Climate Services in Agriculture: A Comprehensive Methodology. *Meteorological Applications, 33*(2), e70170. doi:https://doi.org/10.1002/met.70170

Piani, C., Weedon, G. P., Best, M., Gomes, S. M., Viterbo, P., Hagemann, S., & Haerter, J. O. (2010). Statistical bias correction of global simulated daily precipitation and temperature for the application of hydrological models. *Journal of Hydrology, 395*(3-4), 199-215. doi:https://doi.org/10.1016/j.jhydrol.2010.10.024

Puneet, S., Higinio, R., Ramandeep, B., & Vinay, K. (2023). A new three-step fixed point iteration scheme with strong convergence and applications. *Journal of Computational and Applied Mathematics*, 0377-0427. doi:https://doi.org/10.1016/j.cam.2023.115242

R Core Team. (2021). R: A Language and Environment for Statistical Computing. *R Foundation for Statistical Computing*. Vienna, Austria. Retrieved from https://www.R-project.org/

Rapolaki, R. S., Blamey, R. C., Hermes, J. C., & Reason, C. J. (2021). Moisture sources and transport during an extreme rainfall event over the Limpopo River Basin, southern Africa. *Atmospheric Research, 264*, 105849. doi:https://doi.org/10.1016/j.atmosres.2021.105849

Ross, S. M. (2014). Introduction to probability models. San Diego, CA: Academic Press.

Rummukainen, M. (2010). State-of-the-art with regional climate models. *WIREs Climate Change, 1*(1), 82-96. doi:https://doi.org/10.1002/wcc.8

Rurinda, J., Mapfumo, P., van Wijk, M. T., Mtambanengwe, F., Rufino, M. C., Chikowo, R., & Giller, K. E. (2014). Sources of vulnerability to a variable and changing climate among smallholder households in Zimbabwe: A participatory analysis. *Climate Risk Management, 3*, 65 - 78. doi:https://doi.org/10.1016/j.crm.2014.05.004

Schmidli, J., Frei, C., & Vidale, P. L. (2006). Downscaling from GCM precipitation: a benchmark for dynamical and statistical downscaling methods. *International Journal of Climatology: A Journal of the Royal Meteorological Society, 26*(5), 679-689. doi:https://doi.org/10.1002/joc.1287

Sillmann, J., Thorarinsdottir, T., Keenlyside, N., Schaller, N., Alexander, L., Hegerl, G., . . . Zwiers, F. (2017). Understanding, modeling and predicting weather and climate extremes: Challenges and opportunities. *Weather and Climate Extremes*(18), 65–74.

Soo, E. Z., Jaafar, W. Z., Lai, S. H., Othman, F., Elshafie, A., Islam, T., . . . Hadi, H. S. (2020). Evaluation of bias-adjusted satellite precipitation estimations for extreme flood events in Langat river basin, Malaysia. *Hydrology Research, 51*(1), 105-126. doi:https://doi.org/10.2166/nh.2019.071

Stern, R. D., & Coe, R. (1984). A Model Fitting Analysis of Daily Rainfall Data. *Journal of the Royal Statistical Society. Series A (General), 147*(1), 1–34. doi:https://doi.org/10.2307/2981736

Stern, R. D., & Cooper, P. J. (2011). Assessing Climate Risk and Climate Change using Rainfall Data – A Case Study from Zambia. *Experimental Agriculture, 47*(2), 241-266. doi:https://doi.org/10.1017/S0014479711000081

Teutschbein, C., & Seibert, J. (2012). Bias correction of regional climate model simulations for hydrological climate-change impact studies: Review and evaluation of different methods. *Journal of hydrology, 456*, 12-29. doi:https://doi.org/10.1016/j.jhydrol.2012.05.052

Themeßl, M. J., Gobiet, A., & Leuprecht, A. (2011). Empirical-statistical downscaling and error correction of daily precipitation from regional climate models. *International Journal of Climatology, 31*(10), 1530-1544. doi:https://doi.org/10.1002/joc.2168

Timlin, D., Paff, K., & Han, E. (2024). The role of crop simulation modeling in assessing potential climate change impacts. *Agrosystems, Geosciences & Environment, 7*(1), e20453. doi:https://doi.org/10.1002/agg2.20453

Torgbor, F. F., Stern, D. A., Nkansah, B. K., & Stern, R. D. (2018). Rainfall modelling with a transect view in Ghana. *Ghana Journal of Science, 58*, 41-57. doi:https://doi.org/10.4314/gjs.v58i0

Wilks, D. S. (1999a). Interannual variability and extreme-value characteristics of several stochastic daily precipitation models. *Agricultural and forest meteorology, 93*(3), 153-169. doi:https://doi.org/10.1016/S0168-1923(98)00125-7

Wilks, D. S. (1999b). Multisite downscaling of daily precipitation with a stochastic weather generator. *Climate Research,*, 125–136. doi:https://doi.org/10.3354/cr011125

Wilks, D. S., & Wilby, R. L. (1999). The weather generation game: a review of stochastic weather models. *Progress in physical geography, 23*(3), 329-357. doi:https://doi.org/10.1177/030913339902300302

Williams, C. J., Kniveton, D., & Layberry, R. (2007). Climatic and oceanic associations with daily rainfall extremes over southern Africa. *International Journal of Climatology, 27*(1), 93-108. doi:https://doi.org/10.1002/joc.1376

World Meteorological Organization. (2021). *Guidelines on Surface Station Data Quality Control and Quality Assurance for Climate Applications (WMO-No. 1269).* Geneva, Switzerland: WMO. Retrieved from https://library.wmo.int/idurl/4/57727

Xiao, S., Zou, L., Xia, J., Yang, Z., & Yao, T. (2022). Bias correction framework for satellite precipitation products using a rain/no rain discriminative model. *Science of the Total Environment, 818*, 151679. doi:https://doi.org/10.1016/j.scitotenv.2021.151679

Zambrano-Bigiarini, M., Nauditt, A., Birkel, C., Verbist, K., & Ribbe, L. (2017). Temporal and spatial evaluation of satellite-based rainfall estimates across the complex topographical and climatic gradients of Chile. *Hydrology and Earth System Sciences, 21*(2), 1295-1320. doi:https://doi.org/10.5194/hess-21-1295-2017

# Statements and Declarations

## Funding

This research was funded by a grant from the African Institute for Mathematical Sciences, with financial support from the Government of Canada, provided through Global Affairs Canada, **www.international.gc.ca**, and the International Development Research Centre, **www.idrc.ca**.

## Competing Interests

The authors have no relevant financial or non-financial interests to disclose.

## Author Contributions

Danny Parsons: conceptualisation, methodology, analysis, writing – original draft, writing – review and editing. David Stern: conceptualisation, supervision, writing – review and editing. Mouhamadou Bamba Sylla: supervision, writing – review and editing. James Musyoka: writing – review and editing. John Bagiliko: writing – review and editing. Lily Clements: analysis, writing – review and editing. John Mupuro: data curation, writing – review and editing. Denis Ndanguza: supervision, writing – review and editing. All authors read and approved the final manuscript.

## Data Availability

The AgERA5 reanalysis data used in this study are openly available at the Climate Data Store at https://doi.org/10.24381/cds.6c68c9bb. The station data were provided by the Meteorological Services Department of Zimbabwe and are not publicly available due to data ownership restrictions. The data can be requested from the Meteorological Services Department of Zimbabwe https://www.weatherzw.org.zw/. The corresponding author may also provide the data upon reasonable request, with permission from the Meteorological Services Department of Zimbabwe.

## Code Availability

The code implementing the bias correction methods, together with the scripts used for calibration, validation, and analysis, is available at https://github.com/dannyparsons/markov-chain-bias-correction.